\documentclass{aa}  

\usepackage{graphicx}
\usepackage{txfonts}
\begin{document}

   \title{Photometric Classification of quasars from RCS-2 using Random Forest\thanks{Tables 1, 2 and 3 with the quasar candidates are only available in electronic form
at http://www.aanda.org/}}

%   \subtitle{I. Overviewing the $\kappa$-mechanism}

   \author{D. Carrasco
          \inst{1,2}
          \and
          L. F. Barrientos 
	\inst{1,3}
	\and 
	K. Pichara
	\inst{4,3}
	\and 
	T. Anguita
	\inst{5,3}
	\and D. N. A. Murphy
	\inst{1}
	\and  
	D. G. Gilbank
	\inst{6}
	\and  
	M. D. Gladders
	\inst{7,8}
	\and  
	H. K. C. Yee
	\inst{9}
	\and  
	B. C. Hsieh
	\inst{10}
	\and  
	S. L\'opez
	\inst{11}
          }

   \institute{ Instituto de Astrof\'isica, Pontificia Universidad Cat\'olica de Chile, Avenida Vicu\~na Mackenna 4860, Santiago, Chile
	\and
	School of Physics, University of Melbourne, Parkville, Victoria, Australia
	\email{dcarrasco@student.unimelb.edu.au}
    \and
	Millennium Institute of Astrophysics, Chile
	\and
Departamento de Ciencia de la Computaci\'on, Pontificia Universidad Cat\'olica de Chile, Avenida Vicu\~na Mackenna 4860, Santiago, Chile
	\and
Departamento de Ciencias F\'isicas, Universidad Andres Bello, Avenida Rep\'ublica 252, Santiago, Chile
	\and
South African Astronomical Observatory, P.O. Box 9, Observatory 7935, South Africa
	\and
 Kavli Institute for Cosmological Physics, University of Chicago, 5640 South Ellis Avenue, Chicago, IL 60637, USA
	\and
Department of Astronomy and Astrophysics, University of Chicago, 5640 South Ellis Avenue, Chicago, IL 60637, USA
	\and
Department of Astronomy and Astrophysics, University of Toronto, 60 St. George Street, Toronto, Ontario M5S 3H8, Canada
	\and
Institute of Astrophysics \& Astronomy, Academia Sinica, Taipei 106, Taiwan, R.O.C.
	\and
Departamento de Astronom\'ia, Universidad de Chile, Casilla 36-D, Santiago, Chile
}

%   \date{Received September 15, 1996; accepted March 16, 1997}

% \abstract{}{}{}{}{} 
% 5 {} token are mandatory
 
  \abstract
   {The classification and identification of quasars is fundamental to many astronomical research areas. Given the large volume of photometric survey data available in the near future, automated methods for doing so are required.}
   {Construction of a new quasar candidate catalog from the Red-Sequence Cluster Survey 2 (RCS-2), identified solely from photometric information using an automated algorithm suitable for large surveys. The algorithm performance is tested using a well-defined SDSS spectroscopic sample of quasars and stars.}
   {The Random Forest algorithm constructs the catalog from RCS-2 point sources using SDSS spectroscopically-confirmed stars and quasars. The algorithm identifies putative quasars from broadband magnitudes (g, r, i, z) and colours. Exploiting NUV \textit{GALEX} measurements for a subset of the objects, we refine the classifier by adding new information. An additional subset of the data with WISE W1 and W2 bands is also studied.}
   {Upon analyzing {\bf 542,897} RCS-2 point sources, the algorithm identified {\bf 21,501} quasar candidates, with a training-set-derived \textit{precision} (the fraction of true positives within the group assigned quasar status) of 89.5\% and \textit{recall} (the fraction of true positives relative to all sources that actually are quasars) of 88.4\%. These performance metrics improve for the \textit{GALEX} subset; {\bf 6,530} quasar candidates are identified from {\bf 16,898} sources, with a precision and recall respectively of 97.0\% and 97.5\%. Algorithm performance is further improved when WISE data are included, with \textit{precision} and \textit{recall} increasing to 99.3\% and 99.1\% respectively for {\bf 21,834} quasar candidates from {\bf 242,902} sources. We compile our final catalog ({\bf 38,257}) by merging these samples and removing duplicates. An observational follow up of 17 bright (r < 19) candidates with long-slit spectroscopy at DuPont telescope (LCO) yields 14 confirmed quasars.}
   {The results signal encouraging progress in the classification of point sources with Random Forest algorithms to search for quasars within current and future large-area photometric surveys.}


   \maketitle
%
%________________________________________________________________

\section{Introduction}\label{intro}

Quasars are important astronomical targets, both individually as cosmic lighthouses and within well-defined quasar
catalogues.  As such, their classification and identification becomes
an important, yet non-trivial task.  Their significance in astronomy
has led several groups to search and catalogue them.  These efforts
include: the Large Bright Quasar Survey \citep[LBQS;
  e.g.,][]{foltz89,hewett95}, the FIRST Bright Quasar Survey
\citep[FBQS; e.g.,][]{fbqs1, fbqs2, fbqs3}, the Palomar-Green Survey
of UV-excess Objects \citep{green86}, and the FIRST-2MASS Red Quasar
Survey \citep{glikman2007}.  Among other applications, quasars can be
used to study galaxy evolution \citep[e.g.,][]{hopkins2006}, the
intervening intergalactic gas \citep[e.g.,][]{lopez2008}, cosmological
evolution \citep[e.g.,][]{sqls3}, black hole physics
\citep[e.g.,][]{portinari2012} and the analysis of individual galaxies
and galaxy clusters due to gravitational lensing
\cite[e.g.,][]{faure2009}.  The active nuclei of these galaxies
produce high luminosities (typically $\sim10^{40}$ W) spanning a broad
range of frequencies.  This spectacular luminosity allows them to be observed at high redshifts, which provides an insight
into the distant Universe.  Because of their large distances most
quasars are observed as point sources in optical surveys, meaning they
can easily be misidentified as stellar sources when only photometric
information is available. However, by sampling certain rest-frame
wavelengths, one may distinguish between local and extragalactic
sources through differing spectral characteristics. The lack of a
Balmer jump  (at $\lambda_{\rm rest}$=3646\AA) in low-redshift ($z\lesssim2.2$) quasars separates them from the hot star population.
The Ly$\alpha$ line emission and absorption characterised by the
Ly$\alpha$ forest identified in high-redshift quasar spectra produce
broadband colors that progressively redden with redshift
\citep{richards2002}. Due to observational techniques, most quasar searches are based only on optical
colors \citep[e.g,][]{richards2001,bovy2011}, thus introducing a
redshift-dependent bias. This arises from the distinctive strong line-emission of quasars, impacting their broadband colors relative to the expected continuum flux. Particularly challenging is the selection of quasar targets at intermediate redshifts
($2.2\leqslant z\leqslant3.5$), where classification is typically
inefficient. Quasars with magnitudes brighter than $\sim$21 are
relatively rare, and it can be seen that the quasar and stellar
\textit{loci} cross in color space at $z\sim2.8$
\citep[e.g.,][]{richards2002,bovy2011}.\\

Some previous studies have incorporated multi-wavelength searches by combining different surveys. For example, \citet{richards2006} combine magnitude measurements from different surveys to improve the performance of their quasar selection. Most of the previous work, however, is based on color-color cuts. Whilst effective, the selection and cut limits are both arbitrary and time-consuming, as all possibilities are explored by hand. By contrast, machine learning algorithms have the advantage of adopting, in an automated way, the best criteria to choose quasar candidates based on a sample of objects with pre-defined types. Whilst the creation of such a ``training set'' can be disadvantageous due to the introduction of biases, or indeed due to the compilation of an adequately representative source catalog, this approach permits a fast, efficient classification of big data sets.
\\ 
New methods and approaches
in source classification are required to confront the large volume of
data, due to be taken over the coming years, from next-generation
sky-surveys such as ATLAS \citep{eales2010}, LSST \citep{lsst2009},
DES \citep{des2005} and Pan-STARRS \citep{kaiser2002}. Efficient algorithms are
vital in processing these forthcoming data in order to realise
the science goals of these surveys; an automated
methodology for bulk-classifying the source catalogues will be an
essential ingredient to their success. \\ The main goal of this study
is to use a Machine Learning algorithm to construct a catalog of
quasars selected from purely photometric information. Machine Learning
algorithms have been used to classify objects for many years
\citep[e.g.,][]{ball2006,ball2007,richards2011}. The best known
classification models are: decision trees \citep{quinlan93}, naive
Bayes \citep{duda73}, neural networks \citep{rumelhart86}, Support
Vector Machines \citep{cortes95}, and Random Forest
\citep{breiman2001}. Machine Learning in astronomy, summarized by \citet{ball2010}, has found use in star-galaxy
separation \citep[e.g.,][]{collister2007}, classification of galaxy
morphology \citep[e.g.,][]{huertas2008}, quasar/AGN classification
\citep[e.g.,][]{pichara2013, pichara2012}, galaxy photometric
redshifts \citep[e.g.,][]{gerdes2010} and photometric redshift estimation of quasars
\citep[e.g.,][]{wolf2009}. \\ Our approach for automated quasar identification from only
photometric data is based on the Random Forest algorithm
\citep{breiman2001}, which is a
Machine Learning algorithm based on multiple decision trees
\citep{quinlan86}.  The application of Random Forests in astronomy is
relatively novel, dating back only a few years
\citep[e.g.:][]{dubath2011, richards2011, pichara2012}. A key strength
in the method is an efficient exploration of the spectrum of variable
combinations, whilst avoiding arbitrary thresholding to define
distinct object classes (such as stars, quasars). This approach has been used for the photometric redshift measurement of quasars \citep[e.g.,][]{carliles2010}, whilst \citet{carrasco-kind2013} extend this by also including prediction trees. These techniques are shown to be highly efficient and rapid, although training sets are integral to teaching the model which attributes define an object. Our work is the first to construct a catalog of quasar candidates with the Random Forest algorithm from purely photometric data spanning three different surveys over a wide wavelength range. This multi-wavelength approach has been shown to work well for quasar classification. For example \citet{richards2006} find that adding optical information to MIR data allows for a more efficient type 1 quasar selection. \citet{gao2009} studied the discrimination between quasars and stars with data from catalogs at different wavelengths, concluding that the Random Forest is an effective tool for object classification. Our search is founded primarily on broadband photometric data from
the Red-Sequence Cluster Survey 2 (RCS-2) in addition to supplementary data from {\it GALEX} and WISE surveys. From these data, we
construct a catalog of point sources classified as quasars by the
Random Forest algorithm. This classification prioritizes the precision over the completeness because our aim is to generate a catalog of reliable quasar candidates. \\ The article is organized as follows: in
section \S\ref{data} we present the data used, including RCS-2, SDSS,
WISE, and GALEX, in section \S\ref{rf} we describe the Random Forest
classifier, along with the training and testing sets used. Section
\S\ref{results} describes the results, and in section
\S\ref{summarydiscussion} we present a discussion and summary of our
findings. AB magnitudes are used unless otherwise noted.
\section{Data}\label{data}

In order to construct a point-source training catalog, we cross-match RCS-2 photometric sources with
spectroscopically-confirmed stellar and quasars sources from SDSS. For
subsets of this point-source catalog, we merge WISE and GALEX
photometry. The search radius used in the cross-matching depends on the catalogs used. The source catalogs are described below.

\subsection{RCS-2 data set}\label{rcs2}

The second Red-Sequence Cluster Survey
\citep[RCS-2;][]{gilbank2011} is an optical imaging
survey that aims to detect galaxy clusters in the $0.1\lesssim
z\lesssim1.0$ redshift range. It covers an area of $\sim1,000$
deg$^{2}$. The data were taken at the CFHT telescope using the MegaCam
square-degree imager. The survey was imaged in three filters: {\it g$'$}
(with a 5$\sigma$ point source limiting AB magnitude of 24.4), {\it r$'$}
(limiting magnitude of 24.3), and {\it z$'$} (limiting magnitude of
22.8). The median seeing in the {\it r$'$} band is $0''.71$. About 75\% of
the survey area is also observed in the {\it i$'$} band (limiting magnitude
of 23.7) as part of the Canada-France High-z quasar survey
\citep{willott2005}. Our point-source catalog uses all four bands, which means that the area of search is reduced to $\sim75\%$ ($\sim 750$ deg$^{2}$).

Objects are classified according to their light distribution by
comparing their curve-of-growth with a weighted
average curve derived from a set of four to eight reference
PSFs from nearby unsaturated stars. Each source is categorized by object
type: 0-artefact/spurious object; 1 or 2-galaxy; 3-star; 4-saturated
\citep{yee91,yee96}. All point sources brighter
than $i < 17.5$ are considered as saturated \citep{gilbank2011}. In this study, we selected type-3 (point-source)
objects.

\subsection{SDSS data set}\label{sdss}

The Sloan Digital Sky Survey \citep[SDSS;][]{york2000} is an optical
survey that covers $\sim10,000$ deg$^{2}$ of the sky. The data are
obtained at the Apache Point Observatory, with a dedicated 2.5 meter
telescope and imaged by a large-format mosaic CCD camera. The
optical magnitudes of objects are measured through five optical
broadband filters: u$'$, g$'$, r$'$, i$'$, and z$'$ \citep{fukugita96} with limiting magnitudes of 22.3, 22.6, 22.7, 22.4,
and 20.5 respectively in the AB system. The SDSS PSF is typically
$\sim1''.5$. We use mainly the data from the Data Release Nine \citep[DR9;][]{ahn2012}. It 
is important to clarify that we do not use SDSS magnitudes for the classification.

\subsubsection{Quasars}\label{qsossdss}
Our source quasar catalogue is derived predominantly from DR9. The DR9 Quasar Catalog contains 228,468 quasar
spectra \citep{ahn2012}. It is this quasar sample we cross-match with RCS-2 point sources to obtain a set of spectroscopically confirmed quasars with RCS-2 photometry.

\subsubsection{Stars}\label{starssdss}

Our catalog of stars originates mainly from spectroscopic confirmations of
sources in SDSS Data Release 9. The catalog contains
668,054 confirmed stellar spectra \citep{ahn2012}. We cross-match this combined sample of stars to RCS-2 point source photometry in order to create a catalog of
spectroscopically-confirmed RCS-2 stars.

\subsection{WISE data set}\label{wise}

The Wide-Field Infrared Survey Explorer \citep[WISE;][]{wright2010} is
an infrared all sky survey. It has four mid-IR bands: W1 at 3.4 $\mu
m$, W2 at 4.6 $\mu m$, W3 at 12 $\mu m$, and W4 at 22 $\mu m$ with
angular resolutions of $6''.1$, $6''.4$, $6''.5$, and $10''.1$
respectively. Limiting magnitudes (in Vega) are 16.5 for W1, 15.5 for
W2, 11.2 for W3, and 7.9 for W4; in our study we will use the W1
and W2 bands, following the approach by \citet{stern2012}, as mentioned in $\S$\ref{trs3}. For consistency, we convert these magnitudes to the AB system following \citet{tokunaga2005} and \citet{jarrett2011}.

\subsection{{\it GALEX} data set}\label{galex}

The {\it Galaxy Evolution Explorer} \citep[{\it GALEX};][]{martin2005}
is an orbital space telescope with a mission to compile an all-sky
photometric map in the UV. The telescope images simultaneously with
two bands: the far ultraviolet (FUV; effective wavelength 1528\AA with angular resolution $4''.0$) and the near ultraviolet (NUV; effective wavelength 2271\AA with angular resolution $5''.6$). For the 26,000 deg$^{2}$ All Sky-Imaging (AIS) catalog,
100-second exposures result in limiting (AB)
magnitudes of 19.9 and 20.8 for the FUV and NUV respectively. Whilst
there are other deeper {\it GALEX} catalogs targeting specific
regions, we omit them in preference for a catalog of uniform depth. In
this study, we use the data from GR4/5. \\Because
reddening due to galactic dust becomes significant in the UV, we
correct each {\it GALEX} magnitude with the \citet{schlegel98} dust
maps based on the extinction law from \citet{cardelli89}.

\section{Random Forest}\label{rf}

The Random Forest algorithm \citep{breiman2001} is a tree-based
classification method that learns how to classify objects into
different classes using a training set. In this context, a training set
is a set of pre-classified objects (their class is known); each object is
characterised by a vector whose components are attribute values. To understand how Random Forest classifiers operate, we must first describe their fundamental components: decision trees. A decision
tree \citep{quinlan86} is a graph theory structure where nodes represent attributes
and edges are the possible values the attribute can take. For example, one node
may represent the \lq\lq (g-i) color'' attribute, with two edges pointing out from the node representing two possible values, for example, \lq\lq $\leq 0.5$'' and \lq\lq $> 0.5$''. For a given object, 
depending on the value it has in the attribute \lq\lq (g-i) color'', it will follow a path along one edge or the other, from the node representing the attribute. At the end of the path (after following many other nodes that filter based on the values of the other potential attributes), 
the object will end up in a leaf. Leaves represent a class predicted from the tree: for example, a leaf may
have a value \lq\lq quasar'' or \lq\lq star''. Learning which paths are taken in a decision tree, using well-defined objects, provides us with an automated process to classify unknown objects based on their attributes. The main challenge therefore, is to build a suitable decision tree for a particular
task, in our case, for the automatic classification of quasars/stars. Technical details about the 
building (training) process of a decision tree are out of the scope of this paper, they can be found in \citet{quinlan86}.\\
A Random Forest is an extension of decision trees, with stronger classification capabilities and better performance in many tasks. The core idea of a Random Forest model is to train several decision trees using samples (with replacement) from the training set, and subsequently use these decision trees to classify unknown objects.\\
%The training set is used to build a model describing how the classes
%depend on the values of the vector components.
This model is subsequently applied to a database containing many objects
(with the same attributes) of unknown type in order to make a
prediction of the class they belong to. 
For a training set of $N$ objects described by $F$ attributes, we define $T$ as the number of trees in the Random Forest and
$M < F$ as the number of attributes used in each tree ($T$ and $M$ are model parameters). 
The training procedure is as follows:
\begin{itemize}
\item Generate $T$ data sets with $N$ objects. Each data set is created by randomly sampling objects from the original training set with replacement. This means each of the $T$ sets have the same number of 
elements as the training sets, but some objects are selected more than once.
%This is called \textit{bagging} \citep{raro96}. 
\item From each of the $T$ data sets, grow a full\footnote{A full decision tree means that there is no pruning of the tree during the construction.} decision tree, but on each node select the best split from a set of 
$F$ attributes selected randomly from the $F$ initial attributes. 
%\item Let $M$ be the number of attributes of each object, and $m\ll M$ a constant value during the forest growing such that at each node, $m$ variables are selected randomly from all the $M$ predictor variables. The predictor variable that provides the best split is used to do a binary split in that node. At the next node, it chooses another m variables at random from all predictor variables and do the same. 
%\item Each tree $T$ is grown as much as possible.
\end{itemize}
The creation of each decision tree is both independent and random, and relies on two principles: the first is the diversity among individual classifiers arising from the training of individual trees on different samples. The second principle is that only a subset of randomly selected attributes are used to build each of the trees. As noted in \citet{geurts2006}, these help to find classification
patterns in small subsets of attributes, each tree focusing on different subsets, thus improving the algorithm's accuracy.

Every tree from the forest can assign a class to
an object, based on the attribute values it has. The algorithm's final
predicted classification for a given object is that selected by the
majority of the $T$ trees. Operating in this manner, the Random Forest
algorithm runs efficiently on large databases and can handle
$F\sim10^{3}$ attributes.

To test the classifier we use a 10-fold cross-validation across the training set. This involves partitioning the training set into 10
equal subsets. For a selected subset, we train the model with
the other 9 subsets and test the performance of the resultant
classifications when applied to this selected subset. This procedure
is carried out for each of the 10 subsets. 
%The classifier is trained on all the objects except for one fold.  This held-out set of objects is the test set where the classifier can measure its performance. This is done in turn, holding out each one of the folds to test the classifier on it. 
Results from each of these cross-validated runs are analysed with
performance metrics. To quantify the performance of the algorithm, for
each class of object ({\bf ie. stellar and QSO in this case}) we use
{\it recall} (r), {\it precision} (p), and {\it F-Score} ($\rm{F_{s}}$) \citep{powers}, defined as:\\

   F$_{s}$ $ =  2 \times \displaystyle\frac{{\rm p} \times {\rm r}}{{\rm p + r }}$\\   
   
\noindent where:\\
   
    ${\rm p} = \displaystyle{\rm \frac{t_{p}}{t_{p} + f_{p}} }   \qquad   {\rm r} = {\rm \frac{t_{p}}{t_{p} + f_{n}}}$\\  
  
\noindent $\rm{t_{p}}$, $\rm{f_{p}}$ and $\rm{f_{n}}$ are the number of true positives, false positives and false negatives respectively.

{\it Recall} therefore corresponds to the fraction of correctly classified
objects of each class with respect to all objects genuinely belonging
to that class. \textit{Precision} is the fraction of correctly classified
objects within each class compared to with respect to all objects
classified by the algorithm as members of that class. \textit{F-score} is the
harmonic mean of \textit{precision} and \textit{recall}.\\
The program used for this implementation is the \textit{scikit-learn} \citep{scikit} library for Python.

\subsection{Training Sets}\label{trainingsets}
Training sets are samples of objects for which the target class is
known. In this study, the classes (stellar and QSO sources) are
obtained by cross-identification of SDSS spectroscopically-confirmed
targets to RCS-2 point sources, as discussed above in
$\S$\ref{qsossdss} and $\S$\ref{starssdss}. A match between the two
catalogs is obtained when their angular positions are separated by
less than $0''.5$. The cross-matching is performed for both stellar
and QSO sources, with respectively 20,659 and 8,762 matches made. From
this preliminary matched catalogue we require RCS-2 targets to have
measured flux in all four filters, and photometric errors of less than
0.1 in each. Our catalogs consequently reduce in size to 4,916 quasars
and 10,595 stars. From these data, we create three different training
sets. Each object within the training set is described by attributes
of magnitude and colour. We include all possible attributes to open the parameter space available for the algorithm to ensure an optimal classification.\\ To create the cleanest possible QSO
catalog from the Random Forest algorithm, we select the run with
the highest {\it precision}.\\
To find suitable parameters for the Random Forest, we perform a grid search within a limited discrete space over a set of possible values that will depend on each training set.

\subsubsection{Training Set 1 (TrS1)}\label{trs1}
\begin{figure}[ht!]
\centering
\includegraphics[width=\hsize]{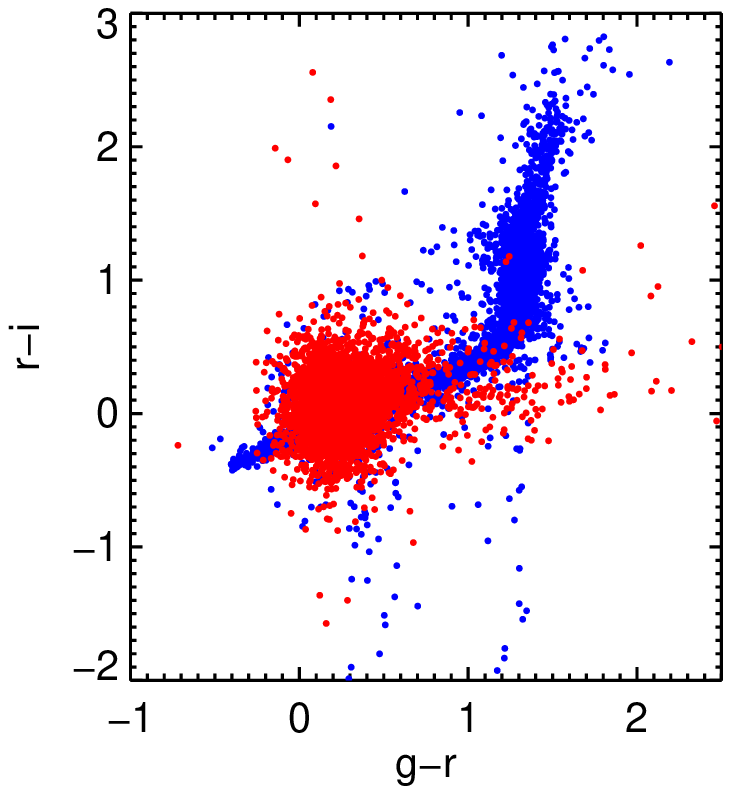}
\includegraphics[width=\hsize]{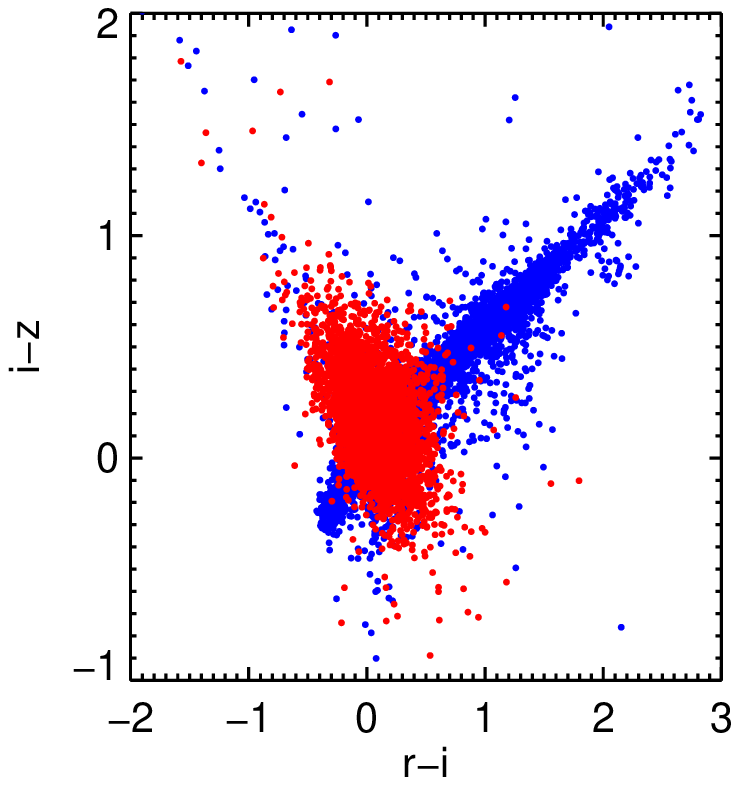}
\caption{Color-color diagrams of spectroscopically confirmed quasars (red) and stars (blue) in RCS-2. The upper diagram shows the colors g-r vs r-i, and the bottom diagram shows the colors r-i vs i-z. Whilst it is possible to see a quasar \textit{clump} and a stellar \textit{locus}, in both cases there is no clean means to separate them. }
\label{sqsdssrcs}
\end{figure}

As discussed above, we create three training sets from the
cross-matched catalog. The attributes used for the first training set are the colors g-r, g-i, g-z, r-i, r-z,
and i-z; the most relevant of these colors are g-r, r-i and i-z. Following
\citet{richards2002}, it is possible to separate quasars and stars
without spectroscopic information by constructing color-color
diagrams. In Figure \ref{sqsdssrcs} we clearly see the characteristic
stellar \textit{locus} and a \textit{clump} of quasars. For the algorithm, as we will see below, it is possible also to use the magnitudes as attributes. We do not do it in this case because the information of the magnitudes will be used in a previous step as we describe in Section \ref{tes1} .

\begin{figure}[ht!]
\includegraphics[width=\hsize]{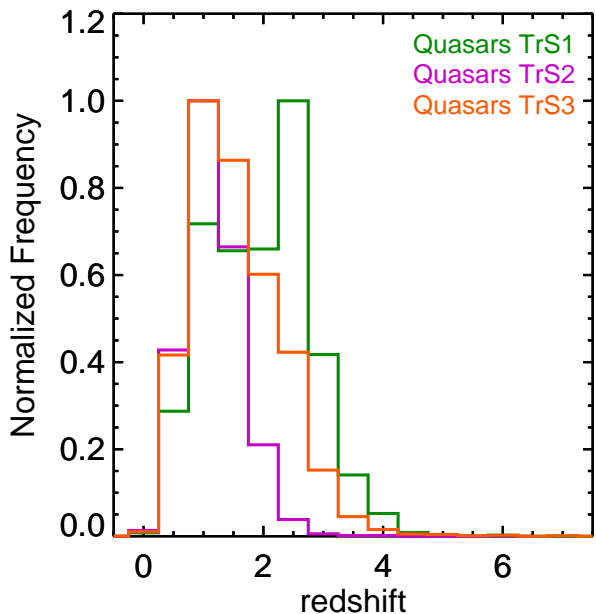}
\caption{The redshift distribution of quasars from the three training sets: TrS1 (green), TrS2 (magenta) and TrS3 (orange). The {\it GALEX}-based data (TrS2) does not probe as deep in redshift as the other two. Whilst the sources with {\it WISE} fluxes (TrS3) appear to peak at a lower redshift, they nevertheless remain sensitive out to the redshifts explored by the optical-only data (TrS1).}
\label{redshifttrs}
\end{figure}

Figure \ref{redshifttrs} shows the large redshift range of
QSOs from this sample in color green: from 0 to 6 and peaking at z$\sim$2.5. In
color-color plots, the quasar population lying near this redshift peak
can be contaminated with stars, making separation of the two
populations more difficult \citep[e.g.:][]{fan99, richards2002,
  bovy2011}.
\begin{figure}[ht!]
\includegraphics[width=\hsize]{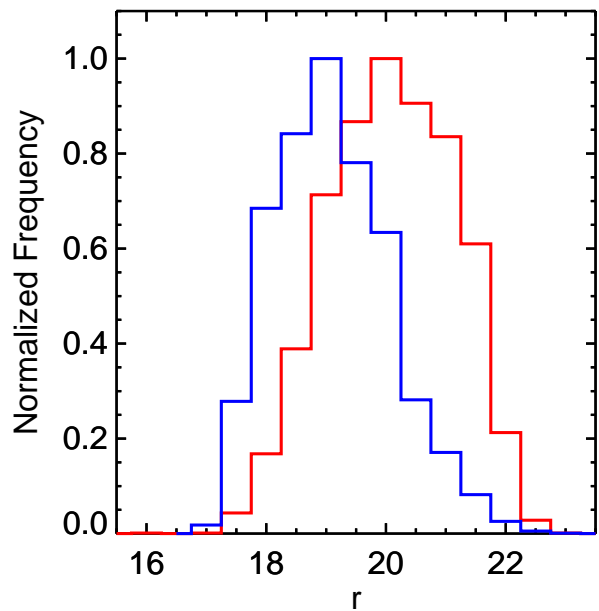}
\caption{The normalized distribution of r-band magnitudes from RCS-2, for quasars and stars listed in the TrS1 training set. We see that the peak of the stellar (blue) sources is approximately one magnitude brighter than that of the quasars (red).}
\label{rmagsdssrcs}
\end{figure}

It is important to analyze magnitude distributions, as fluxes might be
important attributes in distinguishing stars from quasars. Figure
\ref{rmagsdssrcs} therefore shows the magnitude
distribution of stars (blue) and quasars (red), with the former peaking brighter
  than the latter. This means the training set is biased against
faint objects. The number of quasars relative to the number of stars at faint magnitudes is higher; this arises because the fraction of stellar sources drops
off faster than the fraction of QSOs towards fainter
magnitudes. Moreover, the predominance of bright ($r\leq22$) objects
within the training set introduces an observational bias that hinders
accurate classification of faint objects (up to the catalog magnitude
limit of $r\sim23$). To address this excess of quasar classifications (leading to possible misidentification at faint magnitudes) our data set of objects to be classified that matches the magnitude distribution of the training set\footnote{We also approach this problem in a different way to find fainter candidates. The explanation and candidates are available in http://ph.unimelb.edu.au/$\sim$dcarrasco/html/files.html}. The data set is explained in more detail in Section \ref{tes1} .
\\
Our training set TrS1 therefore comprises 4,916 quasars and 10,595
stars.

\subsubsection{Training Set 2 (TrS2)}\label{trs2}

\begin{figure*}[ht!]
\includegraphics{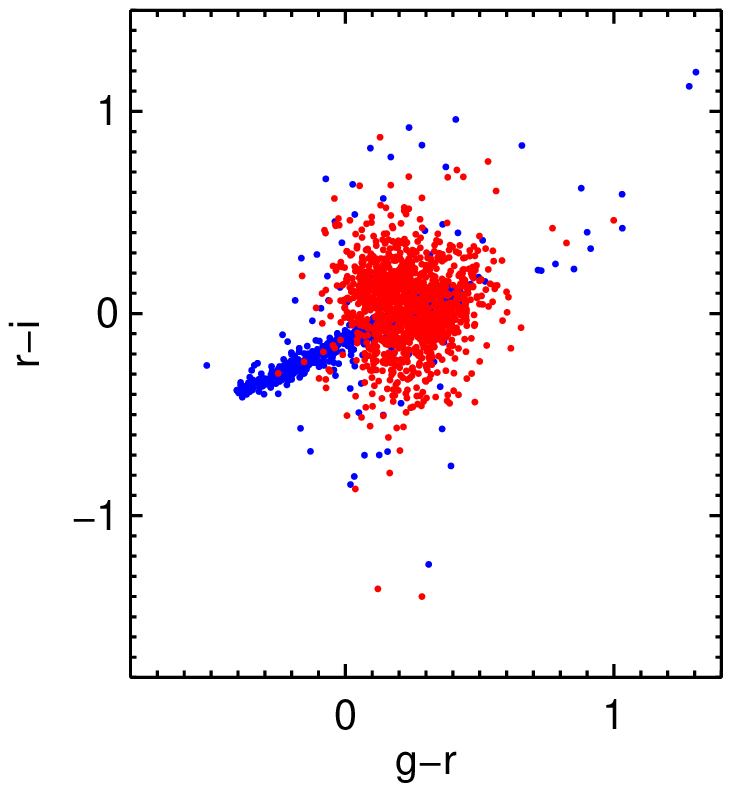}
\includegraphics{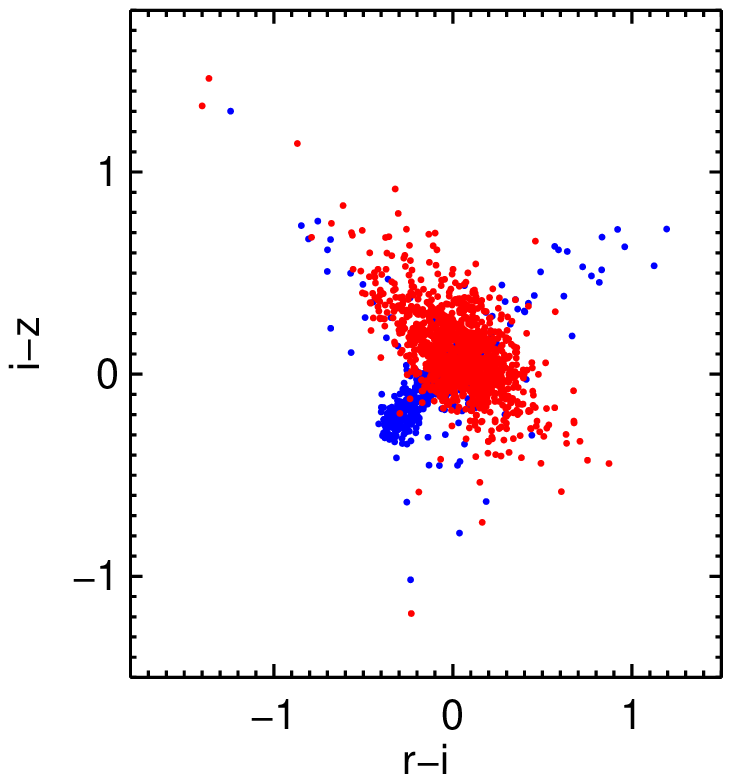}
\centering
\includegraphics{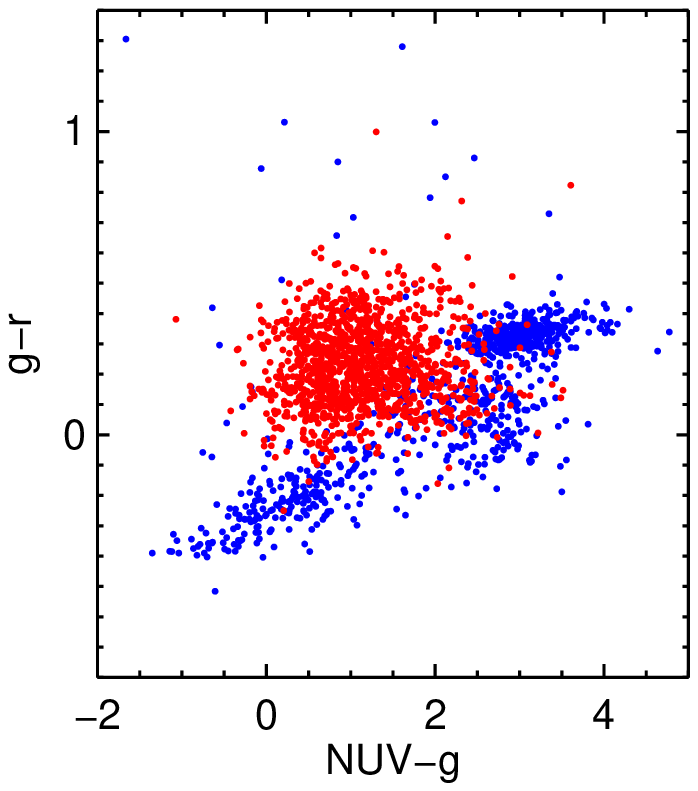}
\caption{Color-color diagrams of spectroscopically confirmed quasars (red) and stars (blue) in RCS-2. The upper diagrams show the optical g-r vs r-i (left) and r-i vs i-z (right) color plots. The bottom plot shows the NUV-g vs g-i color space when including {\it GALEX} data. It can be seen that the inclusion of the UV data provides a clearer separation between quasars and stars compared to solely optical data.}
\label{rcsgalex}
\end{figure*}

\begin{figure}[ht]
\includegraphics[width=\hsize]{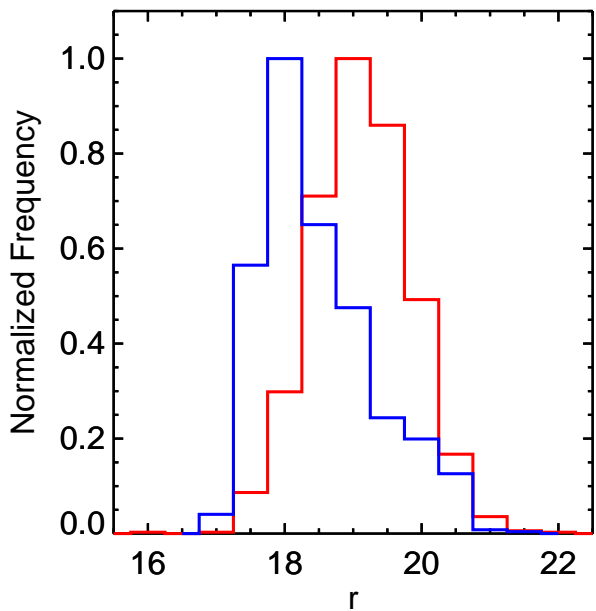}
\caption{Histogram of the normalized r-band magnitude distribution from cross-matched RCS-2 and {\it GALEX} data featured in the TrS2 data-set. Quasars are shown in red, whilst stars are blue. As per the optical data presented in Fig. \ref{rmagsdssrcs}, stellar sources appear to have systematically brighter r-band magnitudes.}
\label{rmagrcsgalex}
\end{figure}

TrS2 is a subset of TrS1. We cross-identify all quasars and stars from
TrS1 with {\it GALEX} objects detected in the NUV band and with a
photometric error of $\leqslant0.25$ (consistent with magnitudes of
$\sim22.5$, around the limit magnitude in this band).  The search
radius used to match sources was $2''.0$, based on the angular
resolution of both surveys. Other studies use larger radii \citep{trammell2007, worseck2011, agueros2005}, but for cross-matching SDSS and GALEX catalogs. It is important to note that whilst our objects were identified from SDSS, we use RCS-2 photometry featuring a better resolution; a smaller search radius is therefore justified. Moreover, unlike the previous works cited before, we do not use the FUV band with a worse resolution. Our cross-matching yields a sample of
1,228 quasars and 815 stars; we note a significant decrease in the
number of objects in the catalog, especially stars. We attribute this
to both the lower angular resolution of {\it GALEX} with respect to
SDSS/RCS-2, and to the redshift dependence of the UV emission from
quasars: only QSOs with redshifts up to $z\sim2.0$ should be
detectable via observed-frame spectral attributes lying within the
filter passband. Because stellar emission in the UV is typically low,
there are considerably fewer stellar sources in this catalog. The stellar sources now present in this new sample will be predominantly blue stars. \\
Inclusion of the {\it GALEX} NUV band is relevant because optical
observations alone do not allow a clean separation between quasars and
stars (See Figure \ref{sqsdssrcs}), especially at intermediate
redshifts ($2.2\lesssim z\lesssim3.5$).  UV flux data are very useful
in quasar classification because stellar-QSO populations are well
separated in UV-optical color space \citep{trammell2007}. In
Figure \ref{rcsgalex} we can see the color-color diagrams of the
quasars and stars with detections in the NUV band. Comparing the NUV-g
vs. g-r plot to optical equivalents, we note the overlap between
quasars and stars has almost disappeared.\\
There is a relation between NUV detection of quasars and redshift. An
important contribution to the bolometric flux is an
intense, broad emission feature dominating the spectral energy
distribution (SED) at bluer wavelengths: the so-called big blue bump \citep{sanders89}. According to \citet{trammell2007}, NUV-band detections of
quasars are almost complete up to $z\sim1.4$, and are still well
recovered at $z\sim1.7$. However, by $z\sim2.0$ the detection
completeness declines to 50\%. While it is not clear whether the FUV
band or NUV band is best suited for quasar detection, we use just NUV
due to the small number ($\sim10\%$) of NUV-detected sources having
FUV fluxes as well. Moreover, the redshift range sampled by FUV
sources appears smaller.  In Figure \ref{redshifttrs}, color magenta shows TrS2
redshift coverage is complete only out to low redshifts compared to
TrS1 (color green). Moreover, the limit of
r-band magnitudes (Figure \ref{rmagrcsgalex}) is much brighter than
TrS1, suggesting the lack of faint-magnitude stars is not a main problem
in this training set. 
For training set TrS2, the attributes used for Random Forest
classification are the four magnitudes from RCS-2 bands: g, r, i, and
z; the NUV band; and the colors: NUV-g, NUV-r, NUV-i, NUV-z, g-r, g-i,
g-z, r-i, r-z, and i-z. As discussed in $\S$\ref{trainingsets}, all
color combinations are added for analysis by the algorithm.

\subsubsection{Training Set 3 (TrS3)}\label{trs3}
\begin{figure*}[ht!]
\centering
\includegraphics{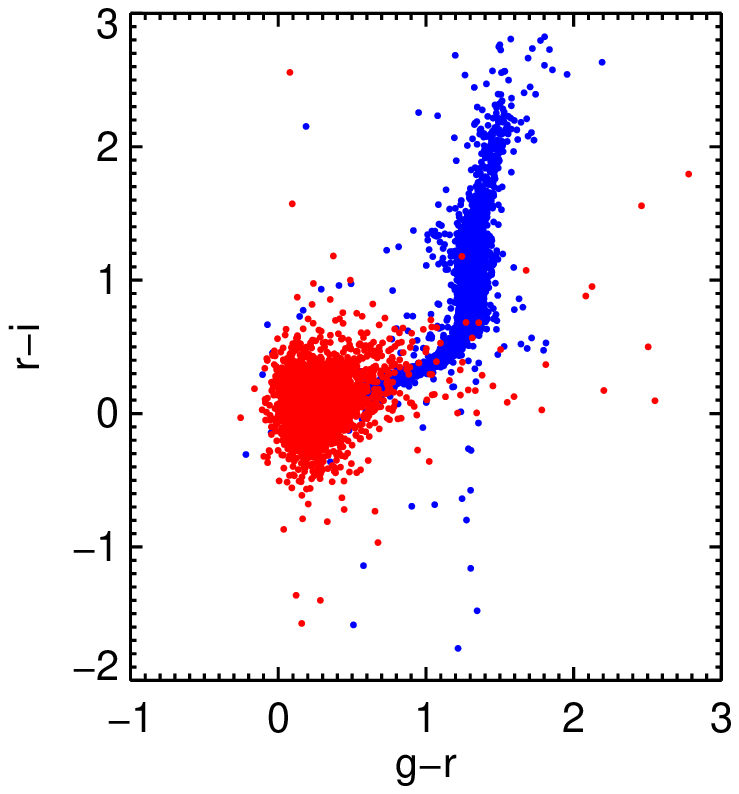}
\includegraphics{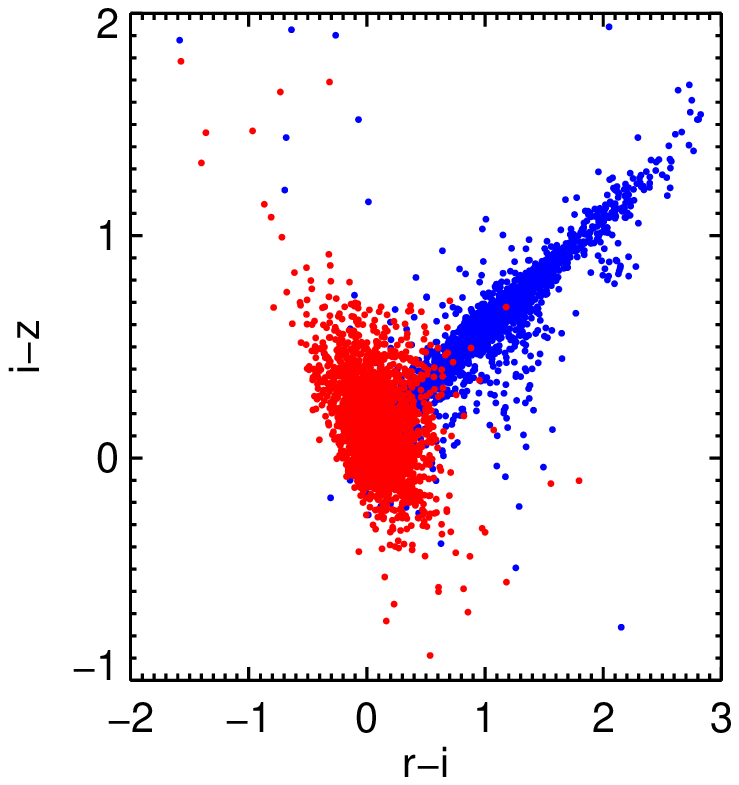}
\includegraphics{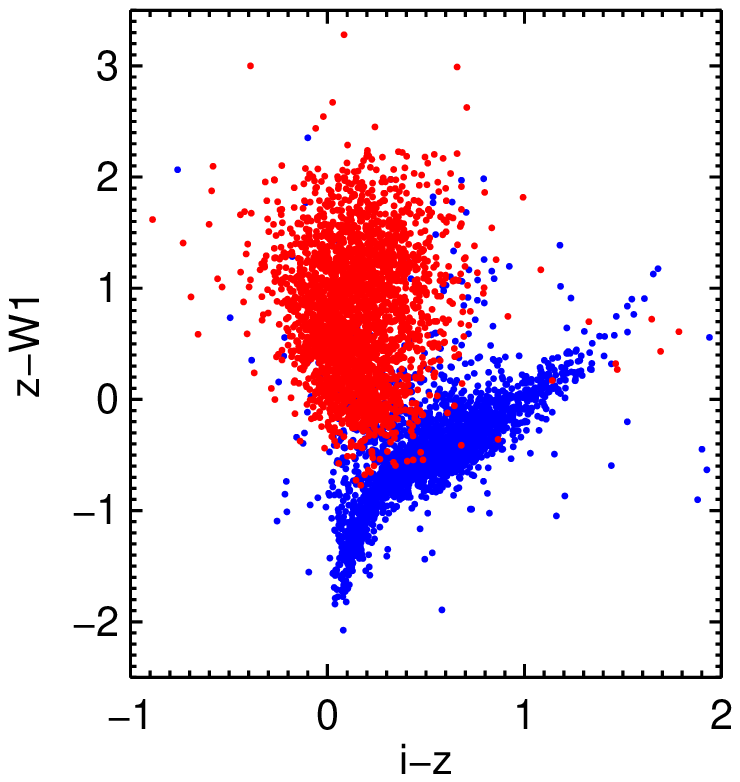}
\includegraphics{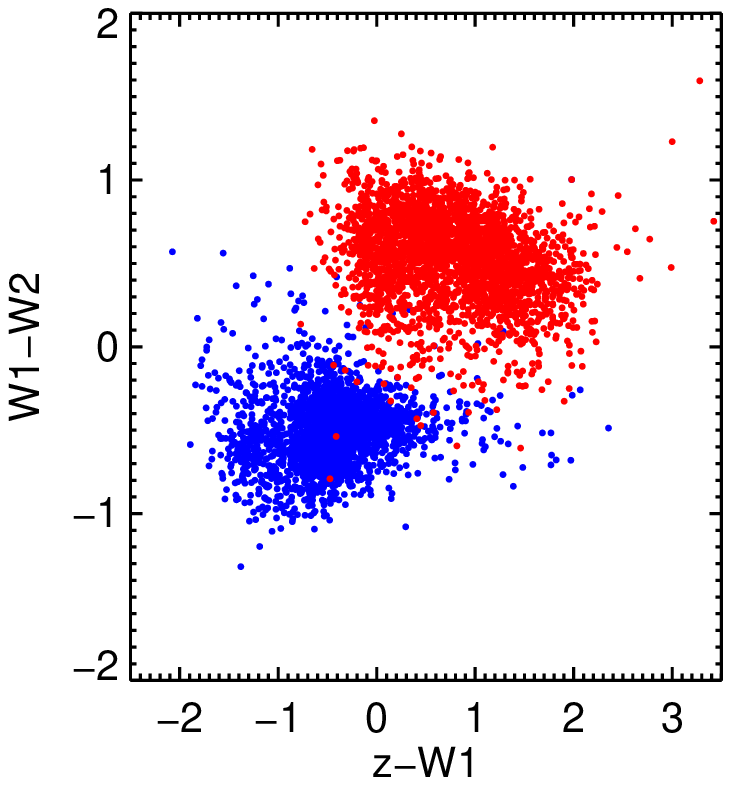}
\caption{Color-color diagrams of the spectroscopically confirmed quasars (red) and stars (blue) in RCS-2 with detection in the W1 and W2 bands from WISE. The upper left diagram shows the colors g-r vs r-i, the upper right diagram shows the colors r-i vs i-z, the bottom left diagram shows the colors i-z vs z-W1, and the bottom left diagram shows the color z-W1 vs W1-W2. In contrast to Figure \ref{sqsdssrcs}, we see the inclusion of these two new bands reduces the overlap between quasars and stars.}
\label{rcswise}
\end{figure*}

This training set is also a subset from TrS1, and is built by
cross-identifying all quasars and stars from TrS1 with WISE sources
detected in the W1 (3.4 $\mu m$) and W2 (4.6 $\mu m$) bands. The
$2''.0$ cross-matching search radius was chosen according to the
angular resolution of both catalogs. Just as in the previous training set, we adopt a smaller matching radius compared to previous studies crossmatching SDSS with WISE sources \citep[e.g.,][]{wu2012,lang2014}. We use these bands following
\citet{stern2012}, where they are used to select quasars from WISE. We
additionally make a cut in the magnitude error corresponding to 0.2 in
both bands, consistent with the magnitude limits of our
sample. Following these selection criteria, we obtain a sample of
2,748 quasars and 2,679 stars. As can be seen in Figure \ref{rcswise},
use of the WISE bands is useful because separation between quasars and
stars in the color-color plots is cleaner than those using purely
optical RCS-2 bands. We expect, therefore, that inclusion of these
bands would boost the performance of the classifier.\\
One additional advantage in using WISE bands, in common with TrS2, is
the introduction of new bands (two in this case) resulting in higher quality
classification. Advantages of TrS3 over TrS2 is the
additional band available to the algorithm but also the wider redshift
coverage, as seen in Figure \ref{redshifttrs} (color orange): redshift coverage
is complete up to $z\sim2$, yet there are still detections up to
$z\sim4$. As such, we are able to classify objects to higher redshifts
than in TrS2. Most significantly, WISE detections cover the
aforementioned mid-redshift range ($2.2\lesssim z\lesssim 3.5$), where
it is hard to separate quasars from stars. For putative QSOs in this
redshift range, the algorithm used in conjunction with TrS3 will be of
great use. Also, as we can see in Figure \ref{rmagrcswise}, the magnitude distribution of this sample, in average, brighter than TrS1 (around 1 magnitude) but fainter than TrS2 (around 0.5 magnitudes), allowing the classification of fainter objects with more information.\\
Attributes of the training set utilised by the algorithm are magnitudes
in the four RCS2 bands: g, r, i, and z; W1 and W2 magnitudes from
WISE; and the colors g-r, g-i, g-z, g-W1, g-W2, r-i, r-z, r-W1, r-W2,
i-z, i-W1, i-W2, z-W1, z-W2, and W1-W2. As explained for the previous
training sets, we make available all colors to the algorithm.

\subsection{Data Set}\label{testset}

A data set is a sample of point sources for which the class is not
known, and where the trained classification model is applied. We have
three data sets, one data set for each one of the training sets
described above. They are all constructed with point sources from
RCS-2 photometry with the same requirements as in the training sets.

\subsubsection{Data Set 1 (DS1)}\label{tes1}

\begin{figure}[ht!]
\centering
\includegraphics[width=\hsize]{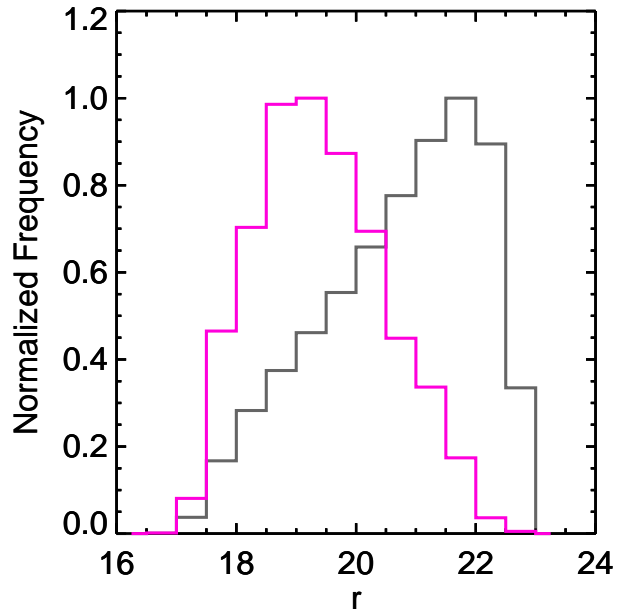}
\caption{Histogram of the normalized magnitude distribution in the r band of the point sources from RCS-2 (color gray) and the new data set DS1 that was obtained by matching its distribution with the distribution of TrS1.}
\label{match_ds1}
\end{figure}

This data set is classified using the algorithm trained on TrS1. It
includes point sources from RCS-2 that meet the requirements described
above. In total, 1,863,970 point sources define what we call data set DS0. However, for consistency we select a subsample from these points matching the magnitude distribution of TrS1. Figure \ref{match_trs1} shows the normalized r magnitude distribution of all the objects (stars and quasars) from TrS1. We take this distribution as the model for our data set. We match the distribution by calculating the fraction of objects in each 0.5 magnitude bin and the number of objects in each bin from the DS0 point source data set. We calculate the number of objects corresponding to the fraction obtained from TrS1 in each bin, randomly selecting this quantity within the same magnitude range. We ultimately end up with a sample of 542,895 photometric sources to be classified by the algorithm. Figure \ref{match_ds1} shows these two distributions: the sample before the cut (DS0, in grey), and the sample after the cut (DS1, in magenta).

\subsubsection{Data Set 2 (DS2)}\label{tes2}

This data set is classified using the algorithm trained on TrS2. It
contains the point sources from DS0 that additionally
have {\it GALEX} NUV-band detections. The cross-matching search radius and photometric error limits are the same than TrS2 as described in Section
\S\ref{trs2}. Within this data set, there are 16,898 sources the
algorithm must classify, of which 9,242 sources are also in the DS1 data set.

\subsubsection{Data Set 3 (DS3)}\label{tes3}

\begin{figure}[h]
\includegraphics{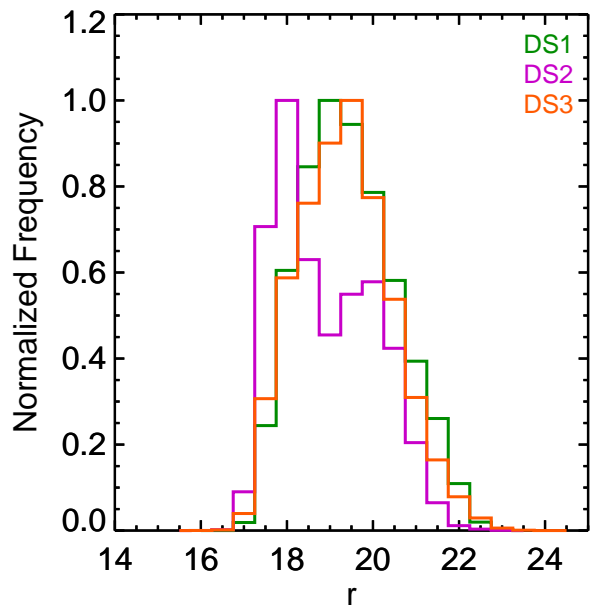}
\caption{The normalized r-band magnitude distribution in RCS-2 for each of our the three test data sets: DS1 (green), DS2 (magenta), DS3 (orange). These data are the sources that the algorithm must classify.}
 \label{rmagds}
\end{figure}

This data set is classified using the algorithm trained on TrS3. The
point sources are those from DS0 with detection in the W1 and W2
bands from WISE.  We apply the TrS3 criteria to the point source
catalog for the cross-match and the limit in photometric errors. These constraints result in a data set of 242,902 point sources for classification. There are 138,658 objects in DS3 that form part of DS1.\\
\\
In Figure \ref{rmagds} we compare the r-band magnitude distribution of the three data sets. As expected, the faintest peak corresponds to DS1 whilst the brightest belongs to DS3. 
\begin{figure}[h]
\includegraphics{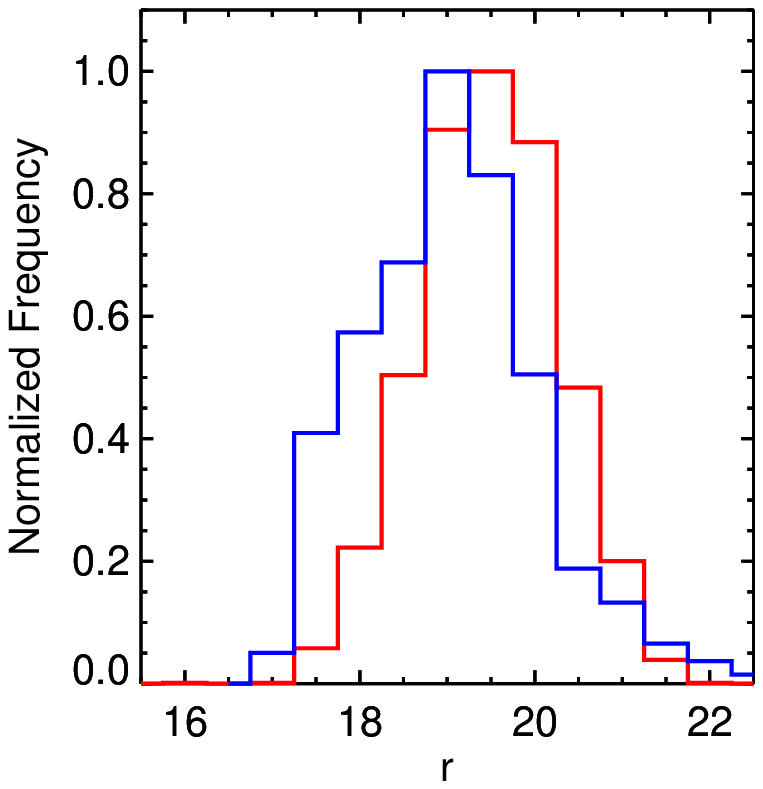}
\caption{Histogram of the normalized magnitude distributions, in the RCS-2 r-band, of quasars and stars from TrS3 (those sources with WISE photometry). As seen previously in Figs. \ref{rmagsdssrcs} and \ref{rmagrcsgalex}, the Quasars (red) are in general fainter than the stars (blue), but in this instance the difference in peak magnitudes is not as pronounced.}
\label{rmagrcswise}
\end{figure} 
\begin{figure}[ht!]
\centering
\includegraphics[width=\hsize]{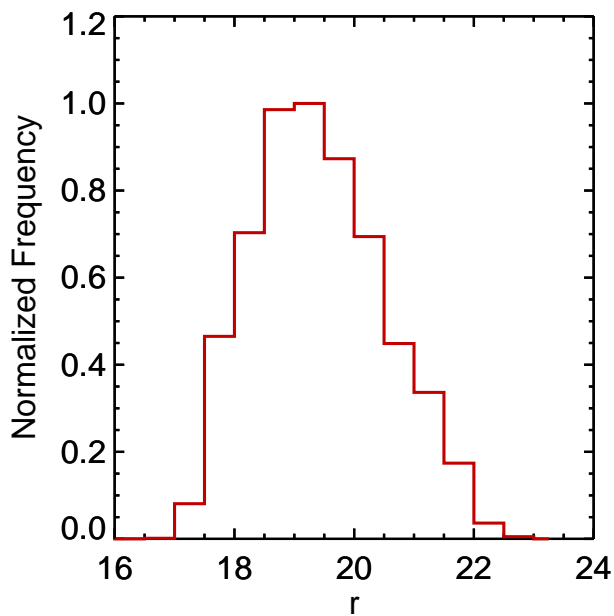}
\caption{Histogram of the normalized magnitude distribution in the r band of all the objects in TrS1.}
\label{match_trs1}
\end{figure}


\section{Results}\label{results}

\begin{figure}[h!]
\includegraphics{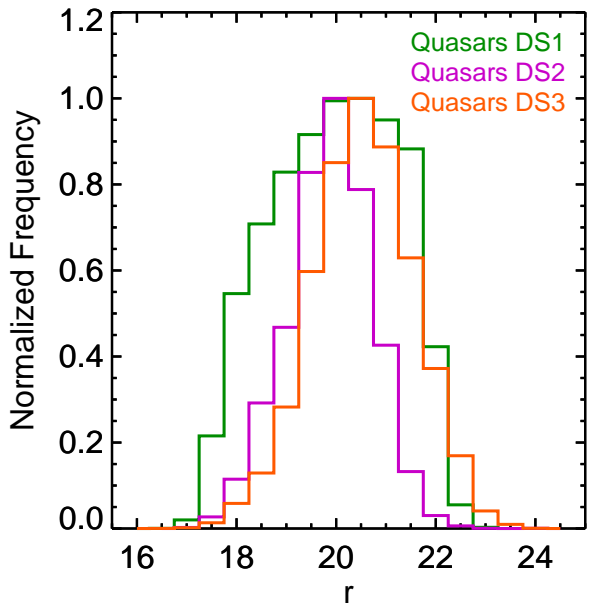}
\caption{The r-band magnitude distribution of quasars classified from the three data sets: DS1 (green), DS2 (magenta) and DS3 (orange).}
\label{rmagteqv3}
\end{figure}

We train the algorithm with the three training sets separately and then apply it to the corresponding data sets:\\
\begin{itemize}

\item For TrS1, our grid search for $M$ (the number of attributes) goes between 3 and 6 with a bin size of 1, and we investigate values of $T$ (the number of trees) between 10 and 150, in bins of 10. The optimal pair of parameters, with an {\it F-score} of 88.9\%, is $\rm{M=4}$ and $\rm{T=50}$. These return {\it recall} and {\it precision} values listed (for both stars and quasars) in the first column of Table \ref{sum2}. We subsequently apply this model to DS1 and obtain 21,501 quasars, which is 4.0\% of the total number of sources. \\
Figure \ref{rmagtes1} shows the r-band magnitude distribution of the two classifications. These distributions are similar to those seen in the TrS1 training set (Figure \ref{rmagsdssrcs}). DS1 stellar classifications (blue) peak at magnitude 19 just as in TrS1, with a similar skew towards brighter magnitudes. The DS1 quasar classifications show a seemingly broader peak (at r$\sim$20.25) than their TrS1 counterparts, with a slight skew towards fainter magnitudes. We note also that the fraction of objects classified as quasars is different between the DS1 candidates ($4.0\%$) and verified objects from TrS1 ($31.7\%$). This is in part due to our prioritization of the precision, but more importantly, it shows that even when the training sample is incomplete, the algorithm is capable of classifying objects that are not covered in the sample (in this case stars). Despite the low fraction of quasars in DS1 relative to TrS1, the fact that their respective r-band distributions for both stellar and quasar classifications are similar is remarkable when considering merely colors are used as attributes for the classification. \\

\begin{figure}[h!]
\includegraphics{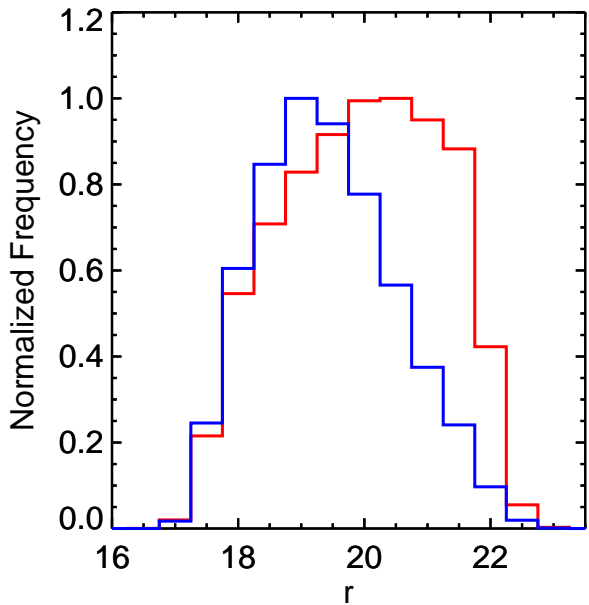}
\caption{Normalized r-band magnitude distributions of from quasar (red) and star (blue) candidates from DS1, as classified by the random forest algorithm. In parallel with the distribution of spectroscopically verified sources presented in Fig. \ref{rmagsdssrcs}, we note that stellar sources are generally brighter than quasars.}
\label{rmagtes1}
\end{figure}

\item For TrS2, our search ranges between 3 and 15 with a bin size of 1 for $M$, and 10 to 150 for $T$. We find that the optimal parameters for source
  classification are $\rm{M=5}$ and $\rm{T=150}$. The {\it recall} and {\it precision} resulting from these parameters are also shown in Table \ref{sum2}, and the {\it F-score} is
  97.2\%. Classifying all sources from DS2, we obtain 6,530 quasars -- corresponding to 38.6\% of the objects.\\

As with DS1, in Figure \ref{rmagtes2} we show the r-band magnitude
distribution for objects classified as stars and QSOs; it can be
seen that the whole sample is brighter than the first one, as
expected. Moreover there is a noted similarity between this
distribution and that of the TrS2 in Figure \ref{rmagrcsgalex}.\\

\begin{figure}[h!]
\includegraphics{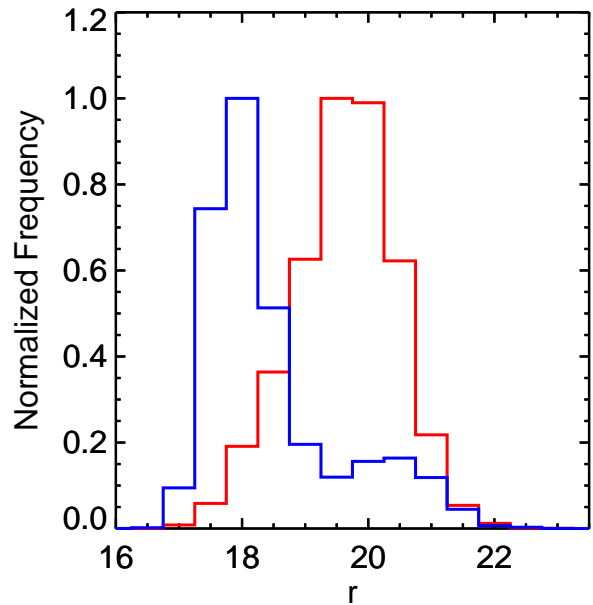}
\caption{Resultant normalized r-band distributions of sources classified, in DS2 by the random forest algorithm, quasars (red) and stars (blue). The difference in these distributions mirrors that presented in the TrS2 training get data shown in Fig. \ref{rmagrcsgalex}.}
\label{rmagtes2}
\end{figure}

\item For TrS3, our search ranges from 3 to 21 for $M$ and from 10 to 150 for $T$. We find the best parameters to be $\rm{M=8}$ and $\rm{T=60}$. The results of {\it recall} and {\it precision} are shown in Table
  \ref{sum2}, and the {\it F-score} is 99.2\%. From the DS3 point sources
  21,834 are classified as quasars, corresponding to a 9.0\% of the
  total. \\
Figure \ref{rmagtes3} shows the r-band magnitude distribution of the
classified objects, and as in the training sets, this sample reaches
fainter magnitudes than DS2, but is brighter than the DS1.\\
\begin{figure}[h!]
\includegraphics{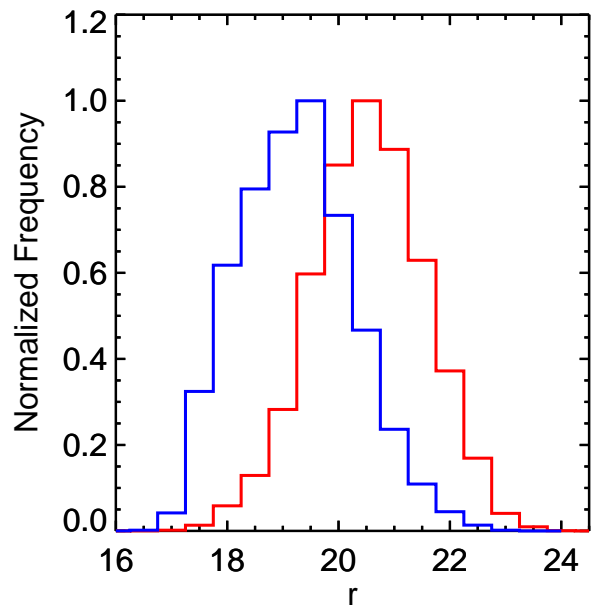}
\caption{The normalized r-band distributions of sources classified as quasars (red) and stars (blue) by the random forest algorithm when applied to the DS3 data set including WISE W1 and W2 photometry. We note there is more separation in these distributions than observed in the TrS3 training set data (Fig. \ref{rmagrcswise}), where the stellar population does not drop off as steeply at fainter magnitudes.}
\label{rmagtes3}
\end{figure}
\end{itemize}
Figure \ref{rmagteqv3} shows the r-band magnitude distribution of quasar candidates from each of the three data sets. The distributions are broadly similar, however we note that quasars from DS1 and DS3 peak at a slightly fainter magnitude ($r=20.5$), than DS2 ($r=20$). This is in accordance to what we expect from the training sets. Furthermore, the DS1 magnitude distribution is broader, relative to TrS1. Quasars from the DS2 have the narrowest distribution of the three data sets, whilst DS3 has the lowest proportion of bright quasars.\\
The total sample of point sources classified was 651,488. Of these 542,897 have detections only in the four optical RCS-2 bands, 16,898 also have {\it GALEX} NUV detections, and 242,902 have W1 and W2-band WISE detections within the stipulated magnitude
error limits. We can assign some degree of confidence to these classifications contingent on how many surveys they have been selected in. Specifically, there are 1,800 objects classified as quasars common to all three data sets. This low number is anticipated
because few objects from the total sample are detected in the NUV
band (the second training set is the smallest one), whilst even fewer
sources have measurements in both WISE and {\it GALEX}. Nevertheless, for this same reason, these objects are very likely to be quasars; having been selected from three different training sets, the reliability of their classification is the strongest. Sources classified as quasars from two of the three data sets comprise 7,514 distinct sources: 1,821 from
DS1 and DS2, 2,788 from DS1 and DS3, and 2,905 from DS2 and DS3). The
remaining category of quasar classifications arise from selection in just one of the three data sets: these consist of 28,943 candidates,
with 14,482 identified from DS1, 120 from DS2, and 14,341 from
DS3. Combining these three groups of classifications, we arrive at 38,257 new quasar candidates from RCS-2. The majority of these arise via
selection from DS1 and DS3, with only a small fraction from DS2. It is important to emphasize that objects only classified as quasars in one of the data sets \textbf{does not} imply they are poor candidates, merely that they \textbf{are not detected in \textit{GALEX} or WISE}, or potentially are selected in just one data set. Inclusion of these extra bands for quasar
classification serves the algorithm well, and allows us to increase
the overall {\it recall} statistic.\\

\begin{sidewaystable*}
\caption{Example of the quasar candidates catalog from DS1.}
\label{cat1}
\begin{center}
\begin{tabular}{ccccccccccccc}
\hline\hline
RA\tablefootmark{a} & DEC\tablefootmark{b} & g\tablefootmark{c} & g$_{\rm err}$\tablefootmark{d} & r\tablefootmark{c} & r$_{\rm err}$\tablefootmark{d} & i\tablefootmark{c} & i$_{\rm err}$\tablefootmark{d} & z\tablefootmark{c} & z$_{\rm err}$\tablefootmark{d} & DS2\tablefootmark{e} & DS3\tablefootmark{e} \\ 
\hline
 0.50114 & +4.70562 & 20.614 & 0.010 & 20.251 & 0.010 & 20.723 & 0.010 & 20.093 & 0.030 & 0 & 0 \\ 
 0.50141 & +2.56122 & 19.186 & 0.000 & 18.840 & 0.000 & 18.855 & 0.030 & 18.968 & 0.100 & 0 & 0 \\
 0.50151 & +2.31300 & 20.319 & 0.010 & 19.785 & 0.000 & 20.043 & 0.100 & 19.705 & 0.010 & 0 & 1 \\
 0.50157 & +2.89213 & 19.456 & 0.000 & 18.847 & 0.020 & 18.534 & 0.000 & 18.592 & 0.070 & 0 & 0 \\
 0.50174 & +3.26450 & 19.926 & 0.000 & 19.569 & 0.000 & 19.753 & 0.010 & 19.673 & 0.010 & 1 & 1 \\
\hline
\end{tabular}
\end{center}
\tablefoot{\\
\tablefoottext{a}{Right Ascension expressed in decimal hours (J2000).}\\
\tablefoottext{b}{Declination expressed in decimal degrees (J2000).}\\
\tablefoottext{c}{AB Magnitudes from RCS-2.}\\
\tablefoottext{d}{Photometric error in the magnitudes from RCS-2.}\\
\tablefoottext{e}{Whether the object is classified as a quasar from the corresponding data set. If true, it is 1. If not, it is 0.}
}
\end{sidewaystable*}

\begin{sidewaystable*}
\caption{Example of the quasar candidates catalog from DS2.\label{cat2}}
\begin{center}
\begin{tabular}{ccccccccccccccc}
\hline\hline
RA\tablefootmark{a} & DEC\tablefootmark{b} & NUV\tablefootmark{c} & NUV$_{\rm err}$\tablefootmark{e} & g\tablefootmark{d} & g$_{\rm err}$\tablefootmark{f} & r\tablefootmark{d} & r$_{\rm err}$\tablefootmark{f} & i\tablefootmark{d} & i$_{\rm err}$\tablefootmark{f} & z\tablefootmark{d} & z$_{\rm err}$\tablefootmark{f} & DS1\tablefootmark{g} & DS2\tablefootmark{g} \\ 
\hline
 0.82091 & -2.67438 & 20.579 &0.136 & 19.932 &0.000 & 19.748 &0.000 &19.460 &0.000 &19.485 &0.010 &0 &1 \\
 0.82116 & -2.50000 & 20.629 &0.138 & 19.526 &0.000 & 19.196 &0.000 &18.903 &0.000 &19.205 &0.010 &0 &1  \\ 
 0.82257 & -2.61910 & 20.724 &0.138 & 19.356 &0.000 & 19.212 &0.000 &19.131 &0.000 &19.012 &0.010 &1 &1 \\
 0.82852 & -2.55673 & 20.888 &0.162 & 19.643 &0.000 & 19.457 &0.000 &20.056 &0.010 &19.605 &0.020 &0 &0 \\
 0.85762 & -2.66545 & 21.061 &0.199 & 20.364 &0.010 & 19.950 &0.010 &20.163 &0.010 &20.008 &0.020 &0 &1 \\
\hline
\end{tabular}
\end{center}
\tablefoot{\\
\tablefoottext{a}{Right Ascension expressed in decimal hours (J2000).}\\
\tablefoottext{b}{Declination expressed in decimal degrees (J2000).}\\
\tablefoottext{c}{AB Magnitudes from GALEX.}\\
\tablefoottext{d}{AB Magnitudes from RCS-2.}\\
\tablefoottext{e}{Photometric error in the magnitudes from GALEX.}\\
\tablefoottext{f}{Photometric error in the magnitudes from RCS-2.}\\
\tablefoottext{g}{Whether the object is classified as a quasar from the corresponding data set. If true, it is 1. If not, it is 0.}
}
\end{sidewaystable*}

\begin{sidewaystable*}
\caption{Example of the quasar candidates catalog from DS3.\label{cat3}}
\begin{center}
\begin{tabular}{ccccccccccccccccc}
\hline\hline
RA\tablefootmark{a} & DEC\tablefootmark{b} & g\tablefootmark{c} & g$_{\rm err}$\tablefootmark{e} & r\tablefootmark{c} & r$_{\rm err}$\tablefootmark{e} & i\tablefootmark{c} & i$_{\rm err}$\tablefootmark{e} & z\tablefootmark{c} & z$_{\rm err}$\tablefootmark{e} & W1\tablefootmark{d} & W1$_{\rm err}$\tablefootmark{f} & W2\tablefootmark{d} & W2$_{\rm err}$\tablefootmark{f} & DS1\tablefootmark{g} & DS3\tablefootmark{g} \\ 
\hline
 0.63326& +4.86742 & 20.324 &0.010 &20.145 &0.010 &20.798 &0.010 &19.977 &0.020 &18.264 &0.056 &18.181 &0.110 &1 &1\\ 
 0.63338& -1.91026 & 20.668 &0.010 &20.352 &0.010 &20.257 &0.010 &19.967 &0.020 &19.459 &0.127 &19.094 &0.195 &1 &0\\
 0.63347& +0.56390 & 20.790 &0.010 &20.361 &0.010 &20.220 &0.010 &20.231 &0.020 &18.983 &0.095 &18.710 &0.153 &0 &1\\ 
 0.63377& +3.78324 & 20.473 &0.010 &20.159 &0.010 &20.051 &0.010 &19.819 &0.020 &18.309 &0.060 &18.051 &0.108 &1 &1  \\ 
 0.63395& -1.33649 & 21.873 &0.030 &21.323 &0.020 &21.409 &0.020 &20.943 &0.050 &19.874 &0.175 &18.906 &0.163 &0
&0  \\
\hline
\end{tabular}
\end{center}
\tablefoot{\\
\tablefoottext{a}{Right Ascension expressed in decimal hours (J2000).}\\
\tablefoottext{b}{Declination expressed in decimal degrees (J2000).}\\
\tablefoottext{c}{AB Magnitudes from RCS-2.}\\
\tablefoottext{d}{AB Magnitudes from WISE}\\
\tablefoottext{e}{Photometric error in the magnitudes from RCS-2.}\\
\tablefoottext{f}{Photometric error in the magnitudes from WISE.}\\
\tablefoottext{g}{Whether the object is classified as a quasar from the corresponding data set. If true, it is 1. If not, it is 0.}
}
\end{sidewaystable*}

For each data set, we construct a
catalog available online (\textit{table1.dat, table2.dat, table3.dat}). In tables \ref{cat1}, \ref{cat2} and \ref{cat3} we show examples of the catalogs. Each contain the coordinates from the RCS-2 survey (in decimal hours), magnitudes in all the bands for which they have been detected (g, r, i, z, NUV, W1, and/or W2), and magnitude errors for these bands. The final two columns indicate whether they are classified as quasars in the other two data sets. As discussed in \S\ref{results}, we stress that candidates classified as quasars in only one of these sets \textbf{does not necessarily signify a lower probability that they are quasars}. Sources may be missed in the other surveys due to either the applied magnitude limit or indeed the redshift of the putative quasar. Nevertheless, this indicator is useful in the case of an object classified as quasar in the three data sets in applying a ``prioritization'' for observational follow-up. For example, with spectroscopic confirmation of only a limited number of targets, those objects selected from all three surveys should be given priority, as they have been selected from three different training sets.\\
In tables \ref{sum1} and \ref{sum2} we summarize the results of the process.\\ 

\begin{figure}[h!]
\includegraphics{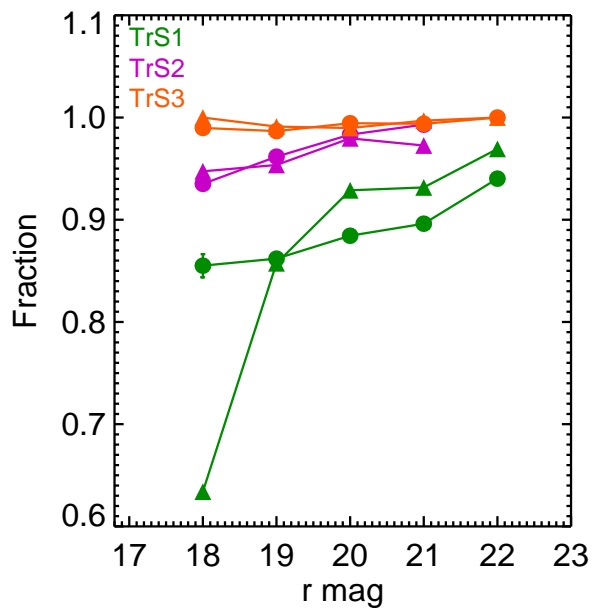}
\caption{Precision (circles) and recall (triangles) for quasars as a function of the r magnitude for the three different training sets: TrS1 (green), TrS2 (magenta) and TrS3 (orange).}
\label{prmag}
\end{figure}

In Figure \ref{prmag} we show the precision and recall for quasars within different magnitude bins, calculated for each of the three training sets with error bars corresponding to the standard error. As discussed previously, these sets have different magnitude limits, meaning there are fewer points sampling TrS3 and TrS2 compared to TrS1. As can be seen, the recall at brighter magnitudes in TrS1 is lower compared to the other samples. We believe this arises from the relative magnitude distributions of stars and galaxies; there are many bright stars compared to quasars such that the algorithm recovers the majority of the stars, but a smaller fraction of quasars. Moreover, missing a few quasars in a small sample produces a larger percentage of missing objects. We attribute the TrS1 increment in the faintest magnitudes to a greater proportion of quasars as seen in Figure \ref{rmagsdssrcs}. For the other cases, both precision and recall are stable with very small variations. 

\subsection{Comparison with XDQSO}

For illustrative purposes, we compare our classifications to a sample drawn from SDSS DR8 XDQSO \citep{bovy2011} study. We randomly select from this catalog a sub sample of QSO or stellar sources with a minimum classifications probability of 70\%. We then match this new sample to each of our independent data sets. The results are: 
\begin{itemize}
\item In DS1, we find 1,450 XDQSO quasar candidates, 82.1\% of which are classified as quasars in our sample. For XDQSO stellar candidates, there are 81,553 objects in our data set, 2.5\% of which are classified as quasars by our algorithm.
\item In DS2, the equivalent match yields 944 XDQSO quasar candidates, of which 97\% are also classified as quasars by our algorithm. We find 1,245 XDQSO star candidates, \textbf{and 9.8\% are classified as quasars in our sample.}
\item In DS3, we find 2,246 XDQSO candidates, with 99.6\% of them among our quasar candidates. For the XDQSO star candidates, we have 34,509 in our data set. 3.0\% of them are classified as quasars in our catalog.
\end{itemize}
We would like to emphasize that these numbers are merely for reference - they do not represent an estimation of the precision or recall of our samples, primarily because sources in the XDQSO catalog are also candidates lacking spectroscopic confirmation; we would need a complete spectroscopic sample of objects with which to compare in order to make a proper assesment. Whilst our study uses a slightly newer release of the SDSS data (DR9 vs. DR8), there is little difference between them in the context of this work, save for some slight astrometric corrections. However, the degree of overlap between the two classification catalogs suggests our results are in good agreement with those drawn from \citet{bovy2011}.

\section{Spectroscopic confirmation}\label{conf}

\begin{figure*}[HT!]
\centering
\includegraphics[scale=0.33,trim = 15mm 0mm 0mm 0mm, clip]{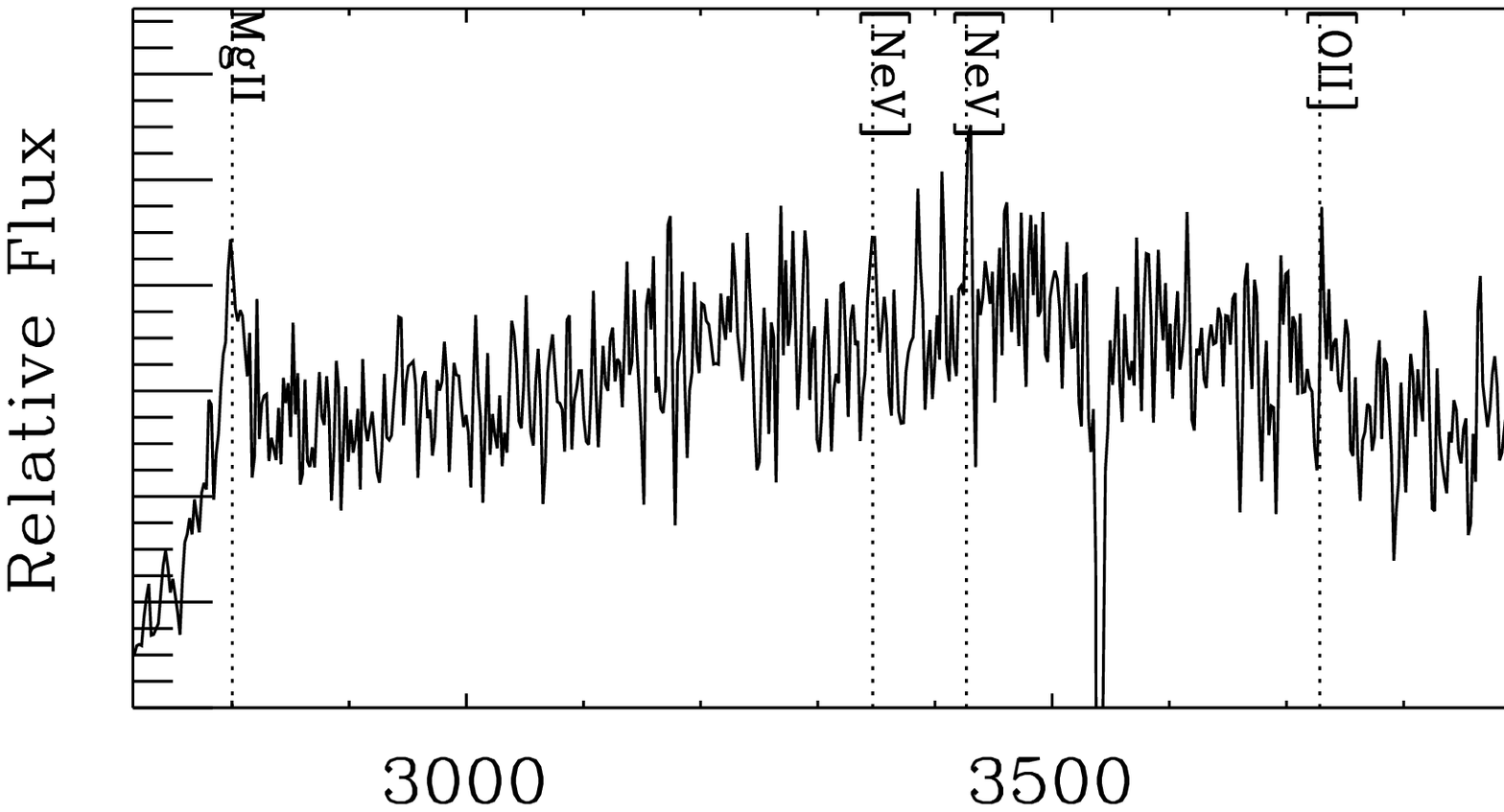}
\caption{Reduced spectra of the confirmed quasars from Table \ref{qsosobs}, taken using the B\&C long-slit spectrograph at the Du Pont 2.5m telescope.\label{spec}}
\end{figure*}

An important step towards validating the putative quasar catalog described in section
\ref{results} is through spectroscopic confirmation of our
candidates. To painstakingly take spectra for all putative QSOs would
be time-consuming and require a long-term observational program outside the scope of the conceptual study we present here. Instead, we randomly sample quasar candidates from the
catalog with $r\leqslant19$ that were classified in at
least 2 of the test sets. This bright magnitude limit was chosen such
that with a modest telescope, observation of many targets was
feasible. Our observations were carried out at the 2.5m DuPont telescope
at Las Campanas Observatory, with the Boller \& Chivens Spectrograph. We
took long-slit spectra of 17 targets covering a wide wavelength range (6230\AA), allowing coverage of the largest possible redshift interval given the instrumental constraints. Whilst this is a
small sample and thus insufficient for detailed statistical analysis, it provides an initial proof-of-concept of a Random Forest algorithm applied to the task of quasar classification. The results of our observations are shown in Table \ref{qsosobs}; reduced spectra of the confirmed quasars and AGN are shown in Figure \ref{spec}.

\begin{table}
\caption{The number of spectroscopically confirmed sources in the TrS1, TrS2 and TrS3 training sets used by the Random Forest algorithm. }
\label{sum1}
\centering
\begin{tabular}{lccc}
\hline\hline
  & TrS1 & TrS2 & TrS3\\
\hline
Number of Quasars& 4,916 & 1,228 & 2,748 \\
Number of Stars& 14,595 & 815 & 2,679 \\
\hline
\end{tabular}
\end{table}

\begin{table}
\caption{Results of the classification of RCS-2 point sources when processed by the Random Forest algorithm. These data have been split tabulated according to their respective data sets, where DS1 comprises solely of optical data, and DS2,3 respectively include {\it GALEX} and {\it WISE} photometry for each source.}
 \label{sum2}
\centering
\begin{tabular}{lccc}
\hline\hline
 & DS1 & DS2 & DS3\\
\hline
N DS & 542,897 & 16,898 & 242,902\\
Precision Q ($\%$) & 89.5 & 97.0 & 99.3 \\
Recall Q ($\%$) &  88.4 & 97.5 & 99.1 \\
Precision S\ ($\%$) & 94.8 & 96.4 & 99.2 \\
Recall S ($\%$) & 95.0 & 95.2 & 99.1 \\
N Quasars & 21,501 & 6,530 & 21,834 \\
\% Quasars & 4.0 & 38.6 & 9.0\\
\hline
\end{tabular}
\\
\tablefoot{\textit{N DS} is the number of objects from the data set. \textit{Precision Q}, \textit{Recall Q}, \textit{Precision S}, and \textit{Recall Q} are the {\it precision} for quasars, {\it recall} for quasars, {\it precision} for stars, and {\it recall} for stars, respectively. \textit{N Quasars} is the number of objects classified as quasars from the data set, and \textit{\% Quasars} is the percentage of those classified quasars from all the objects of the data set.}
\end{table}
\begin{table*}
\caption{Overview of our spectroscopic analysis of 17 quasar candidates selected by the random forest algorithm, and followed-up with long-slit spectroscopy at the Du Pont telescope. We note, in columns 3-5, where a target was classified as a quasar in the respective data sets. Only one star is amongst this sample, along with one galaxy not classed as active. Reduced spectra of the quasars can be found in Fig. \ref{spec}.}
\label{qsosobs}
\begin{center}
\begin{tabular}{ccccccc}
\hline\hline
Object\tablefootmark{a} & r mag\tablefootmark{b} & DS1\tablefootmark{c} & DS2\tablefootmark{c} & DS3\tablefootmark{c} & Spec\tablefootmark{d} & Redshift\tablefootmark{e}\\
\hline
RCS2 1243-0438 & 17.45 & 1 & 1 & 1 & Quasar & 0.79 \\
RCS2 1303-0507 & 17.90 & 1 & 1 & 1 & Quasar & 0.91 \\
RCS2 1307-0439 & 18.01 & 1 & 0 & 1 & Quasar & 1.16 \\
RCS2 1312-0447 & 17.75 & 1 & 1 & 1 & Quasar & 0.53 \\
RCS2 1250-0443 & 18.15 & 1 & 1 & 1 & Quasar & 1.80 \\
RCS2 1251-0435 & 18.20 & 1 & 1 & 1 & Quasar & 1.63 \\
RCS2 1252-0524 & 18.00 & 1 & 0 & 1 & Quasar & 2.26 \\
RCS2 1301-0518 & 18.51 & 1 & 1 & 1 & Quasar & 0.55 \\
RCS2 1303-0448 & 18.53 & 1 & 0 & 1 & Quasar & 2.36 \\
RCS2 1304-0456 & 18.34 & 1 & 1 & 1 & Quasar & 0.76 \\
RCS2 1057-0448 & 17.62 & 1 & 1 & 1 & Quasar & 0.69 \\
RCS2 1100-0313 & 18.18 & 1 & 1 & 1 & Star & - \\
RCS2 1101-0846 & 18.18 & 1 & 1 & 1 & AGN & 0.39 \\
RCS2 1106-0821 & 18.19 & 1 & 1 & 1 & Quasar & 1.43 \\
RCS2 1310-0458 & 18.40 & 1 & 0 & 1 & Quasar & 2.65 \\
RCS2 1303-0505 & 18.64 & 1 & 0 & 1 & Quasar & 0.48 \\
RCS2 1305-0435 & 18.48 & 1 & 1 & 1 & Galaxy & - \\ 
\hline
\end{tabular}
\end{center}
\tablefoot{\\
\tablefoottext{a}{Object is the RCS-2 name.}\\
\tablefoottext{b}{RCS-2 r band magnitude.}\\
\tablefoottext{c}{Whether the object is classified as a quasar from the corresponding data set. If true, it is 1. If not, it is 0.}\\
\tablefoottext{d}{Spectroscopic classification of the object.}\\
\tablefoottext{e}{Spectroscopic redshift of the object.}
}
\end{table*}

We find 14 of the 17 candidates can be confirmed as quasars. Of the remaining three, one could also be considered an AGN with narrower emission lines. We note that AGN such as this target are expected to have similar colors to those of quasars. It is significant that the two false-positives were classified by the random forest algorithm based on RCS-2, NUV and WISE data, further reinforcing that joint classification by these three datasets is not a ``gold-standard'' for successful classification. Indeed, 5 of the 17 targets (or 36\% of the quasars) did not fall under this category yet were verified as quasars.

Whilst our small sample size precludes detailed statistical insight, we nevertheless consider our results satisfactory. In the bright regime we sampled spectroscopically, there are a higher fraction of stars in the training sets. As we illustrate in Figure \ref{prmag}, the precision in this region is characteristically lower compared to fainter magnitudes. Therefore, under the conditions of a complete training set, the algorithm performs very well. We highlight the identification of 3 QSOs between $2.2\lesssim z\lesssim3.5$: it is in this regime that stellar and quasar populations typically overlap in optical color-color space.

\section{Summary and Discussion}\label{summarydiscussion}

Based on a Random Forest algorithm, we built a catalog containing
38,257 new quasar candidates, with a {\it precision}
over $\sim$90\%. A subset of these, 24\% of the catalog, have
photometric detections in {\it GALEX} and/or WISE) and accordingly
achieve {\it precisions} of at least $\sim97\%$. The increase in {\it precision}
with additional bands is anticipated, as the Random Forest algorithm
performance improves when more information is provided. This is only
significant, however, when that information assists in the separation
of the object classes: the additional bands in this 24\% subset greatly assist in
separating quasars from stars in color-color space, as seen in Figures
\ref{rmagrcsgalex} and \ref{rmagrcswise}). These results are comparable to those from different quasar candidate searches. For example, \citet{richards2009} find an overall efficiency (similar to \textit{recall} in our terms) of $\sim80\%$, rising to over $97\%$ when UVX information is added.

Having been trained with a catalog of spectroscopically-confirmed stars and quasars, the Random Forest algorithm is applied to RCS-2 point sources. We require that these point sources have
measured magnitudes in each of the four RCS-2 bands with photometric errors below 0.1
magnitudes. From these point sources, three data sets (DS1-3) are
compiled. We first construct a data set (DS0) with all point sources from RCS-2 with
the aforementioned requirements. The first data set we use for classification (DS1) is a random subset of DS0 matching the r-band magnitude distribution of the training set (Figure \ref{match_trs1}). The second (DS2), another subset of DS0,
contains RCS-2 point sources with NUV-band {\it GALEX} detections. The
third (DS3), also a subset of DS0, contains RCS-2 point sources with
detections in the W1 and W2 WISE bands. We construct quasar candidate catalogs
for each of these data sets. The first, with 21,501 quasar
classifications from a total of 542,897 point sources, corresponds
to 4.0 \% of the sample. DS2 has 6,530 quasar candidates from 16,898
point sources (38.6\% of the total sample), whilst DS3 selects 21,834 quasar
candidates from 242,902 point sources (corresponding to 9.0\% of the
data set). Merging these quasar classifications and
removing duplicates provides us with our final sample of 38,257 quasar
candidates. This catalog is split into three (\textit{table1.dat, table2.dat, table3.dat}), and is available at the CDS: they contain the coordinates from RCS-2 survey,
magnitude in all detected bands (NUV, g, r, i, z, W1, W2), photometric
errors of those magnitudes, and the data sets where they were classified as quasars.\\
The number of data sets within which a candidate was classified as a quasar does not indicate the reliability of the classification, merely that they (for example)
have no WISE or {\it GALEX} data. Nevertheless, this is a
useful indicator when seeking candidates that have the highest
probability of being genuine quasars for the purposes of spectroscopic follow-ups.\\ We have obtained a reliable new sample of quasars candidates that can be used for a wide range of
astronomical applications. From this sample, we obtained the spectra of 17 candidates with
magnitudes $r\sim19$, that were classified as quasars in at least 2 data sets. From this small sample, 14 were confirmed as quasars, 1 as an AGN. Even within this small sample, there is good
agreement with our expectations.\\ The Random Forest algorithm works well in the
classification of point sources into quasars and stars based on
magnitude and color information. The algorithm is useful because it
automatically chooses the attributes optimally separate the two classes of object. The approach described here has broad applicability, permitting a
similar studies on future photometric surveys such as LSST, requiring only a
training set and photometric information. The advantage of the Random
Forest over many other approaches is its high level of automatization
and suitability for processing large volumes of data. Training sets
need be neither new or large - our sample of spectroscopically confirmed SDSS
sources, cross-matched with RCS-2 point sources, were entirely suitable for our purposes in this
study. Extensions to the work we detail could entail the inclusion of additional photometric information from other all-sky surveys, an application of the existing algorithm and training sets to new, larger
photometric catalogs, and spectroscopic follow-up of Random Forest quasar candidates in order to gain an insight into the performance of the technique beyond that explored here.\\
\begin{figure*}[HT!]
\centering
\includegraphics[scale=0.33,trim = 15mm 15mm 0mm 0mm, clip]{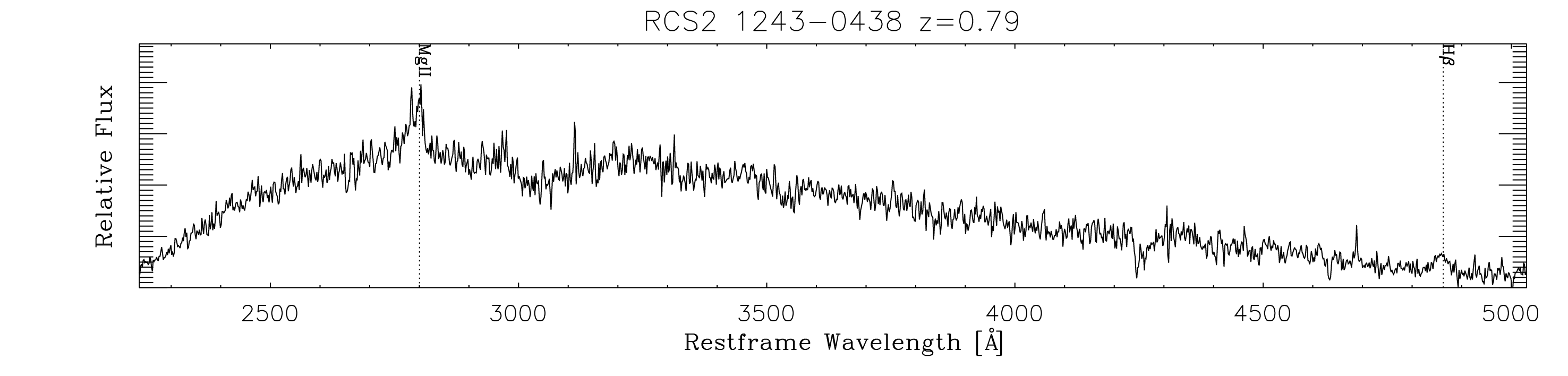}
\includegraphics[scale=0.33,trim = 15mm 15mm 0mm 0mm, clip]{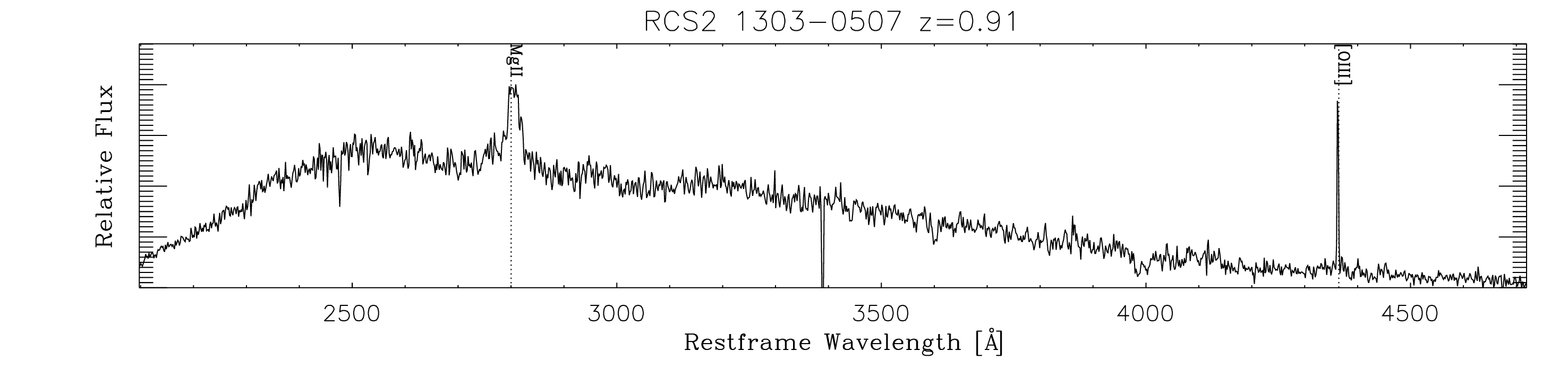}
\includegraphics[scale=0.33,trim = 15mm 15mm 0mm 0mm, clip]{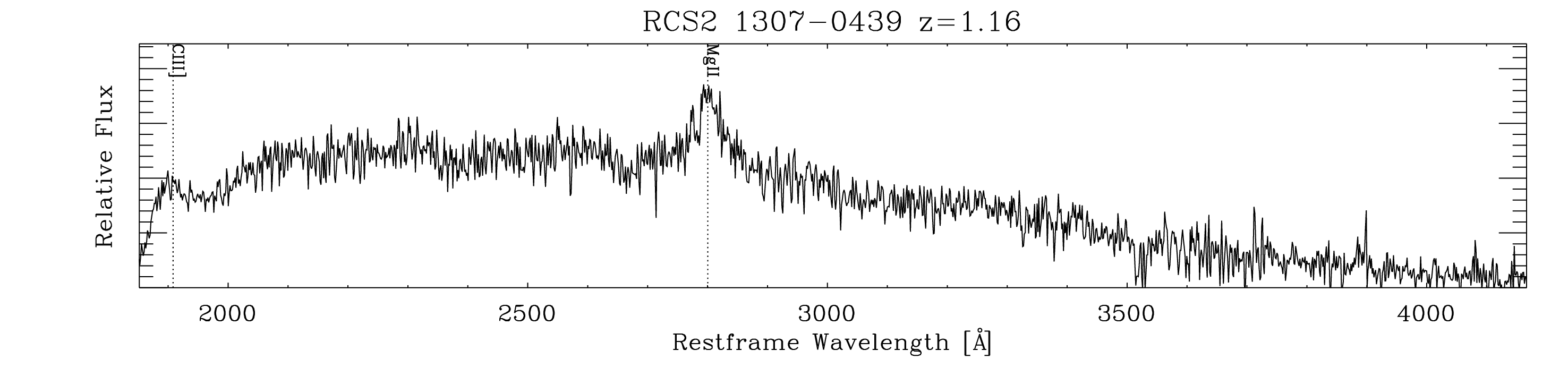}
\includegraphics[scale=0.33,trim = 15mm 15mm 0mm 0mm, clip]{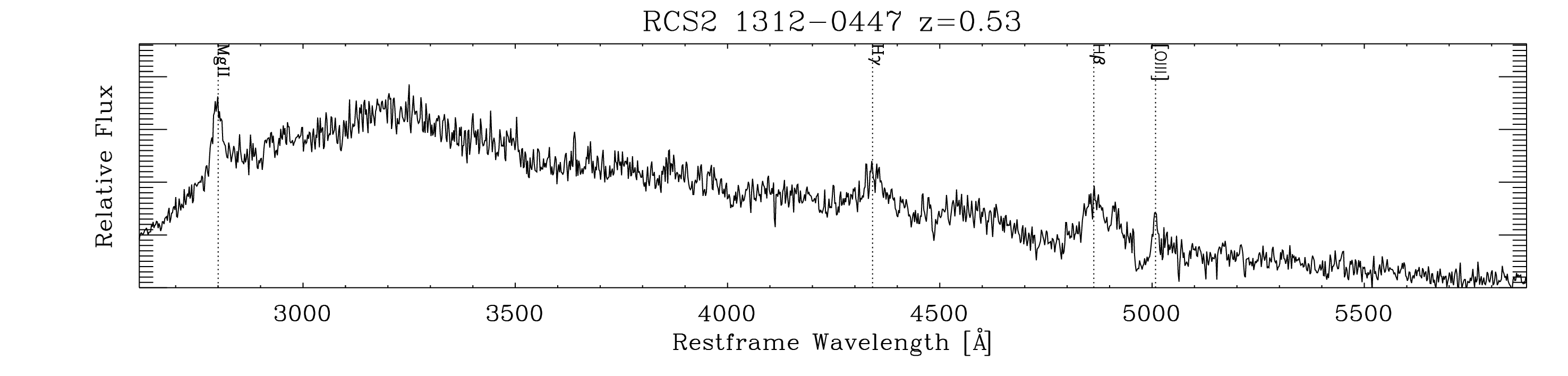}
\includegraphics[scale=0.33,trim = 15mm 15mm 0mm 0mm, clip]{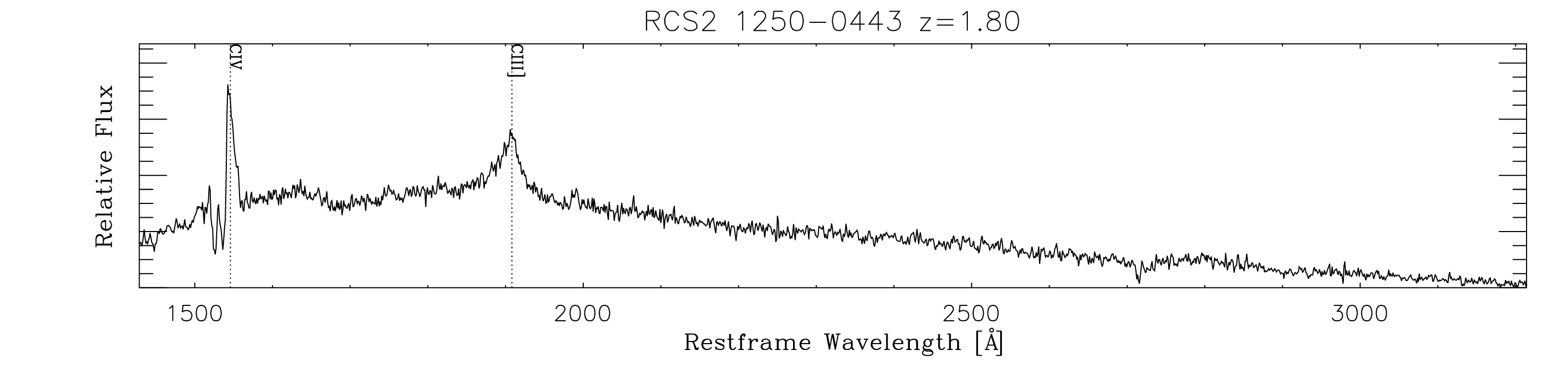}
\includegraphics[scale=0.33,trim = 15mm 15mm 0mm 0mm, clip]{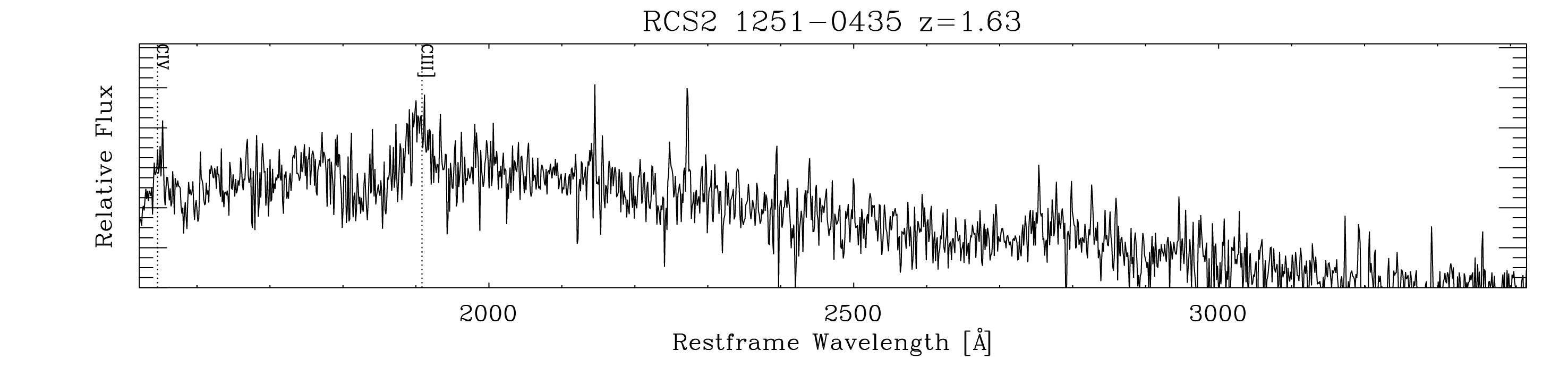}
\includegraphics[scale=0.33,trim = 15mm 0mm 0mm 0mm, clip]{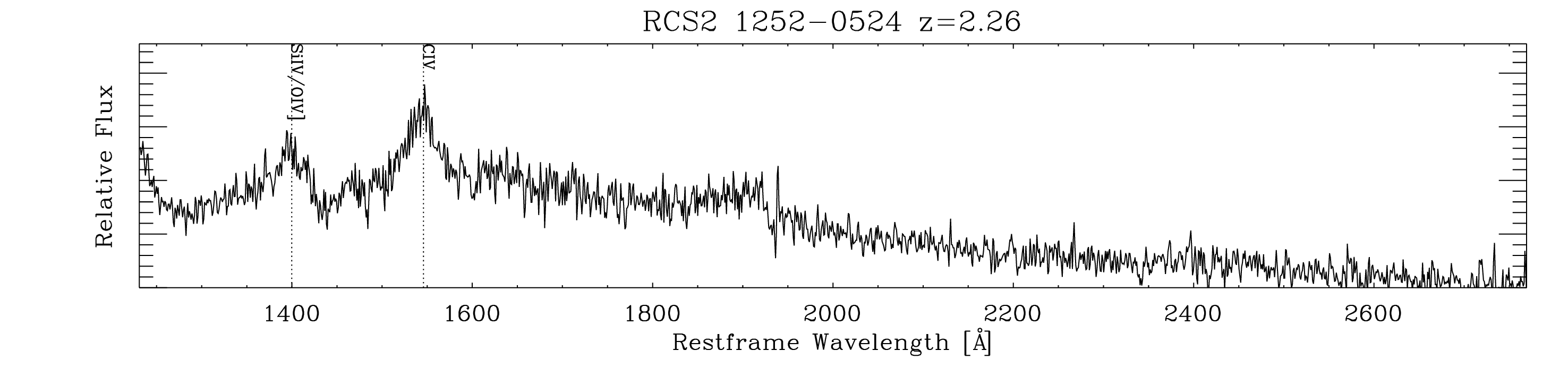}
\end{figure*}
\begin{figure*}[HT!]
\centering
\includegraphics[scale=0.33,trim = 15mm 15mm 0mm 0mm, clip]{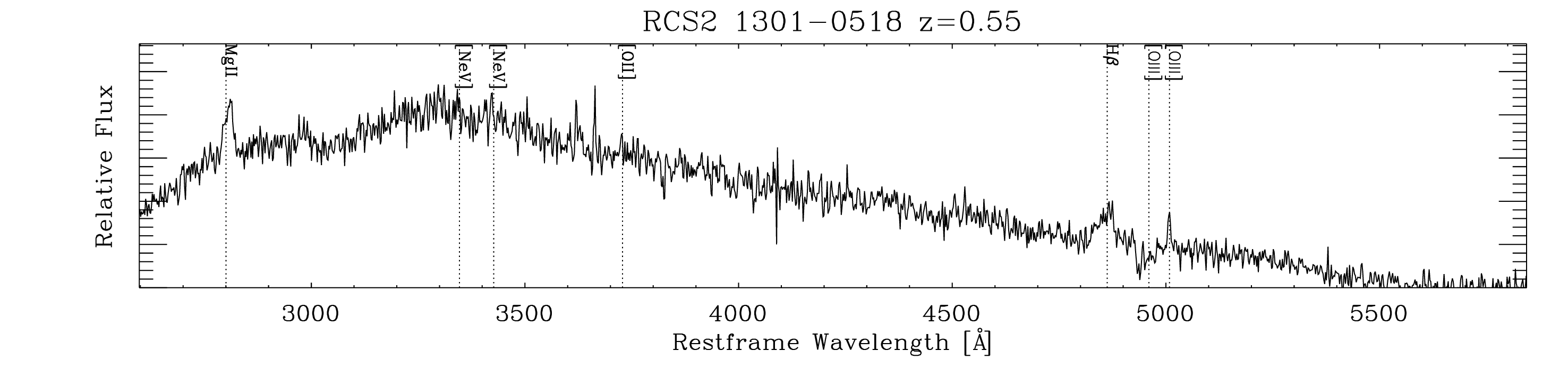}
\includegraphics[scale=0.33,trim = 15mm 15mm 0mm 0mm, clip]{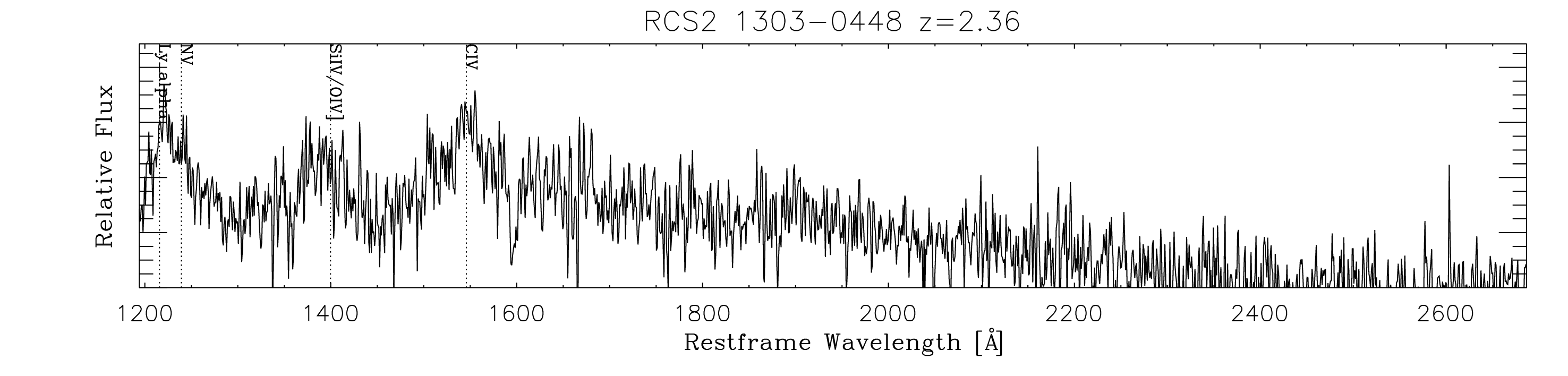}
\includegraphics[scale=0.33,trim = 15mm 15mm 0mm 0mm, clip]{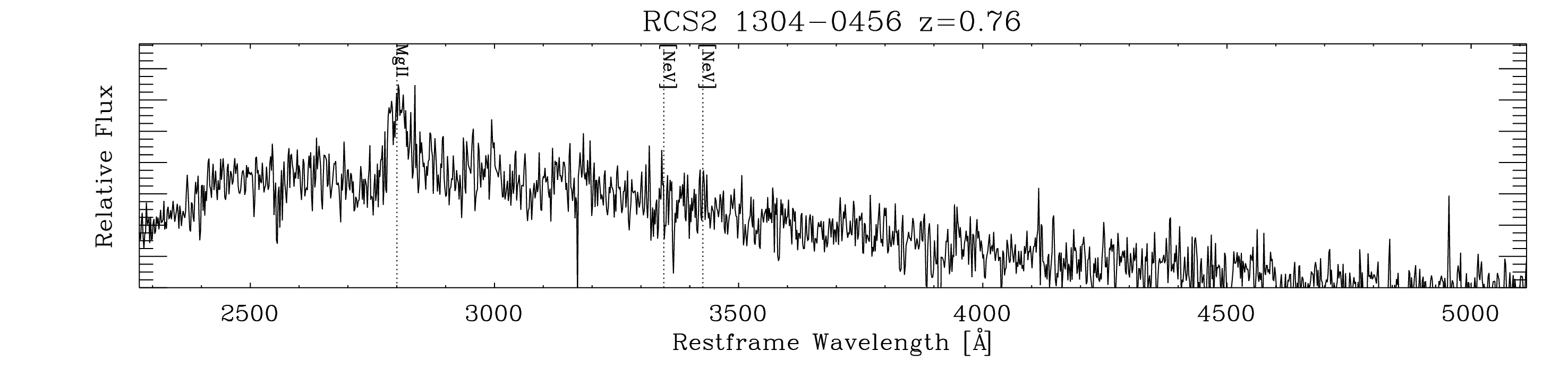}
\includegraphics[scale=0.33,trim = 15mm 15mm 0mm 0mm, clip]{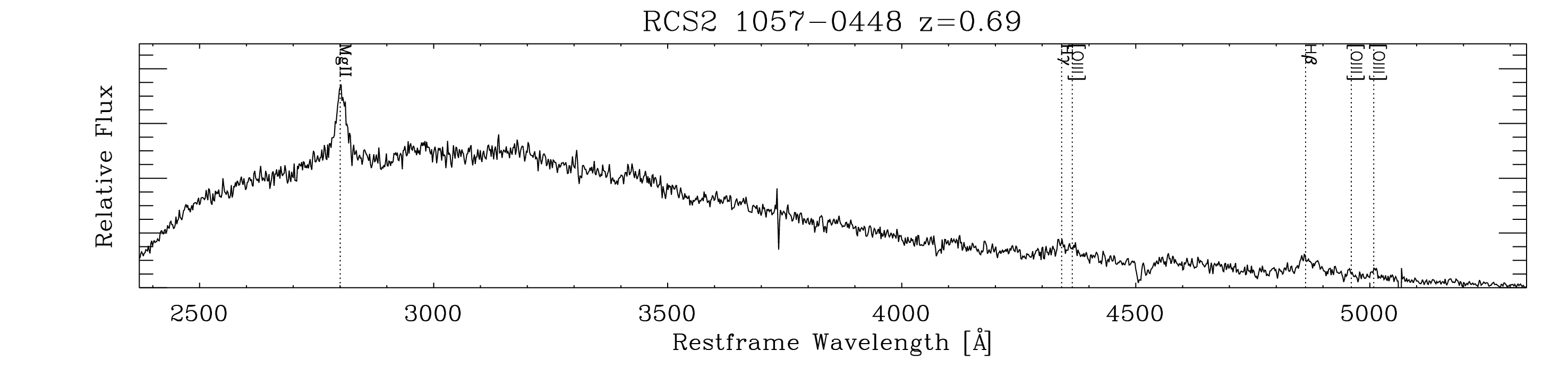}
\includegraphics[scale=0.33,trim = 15mm 15mm 0mm 0mm, clip]{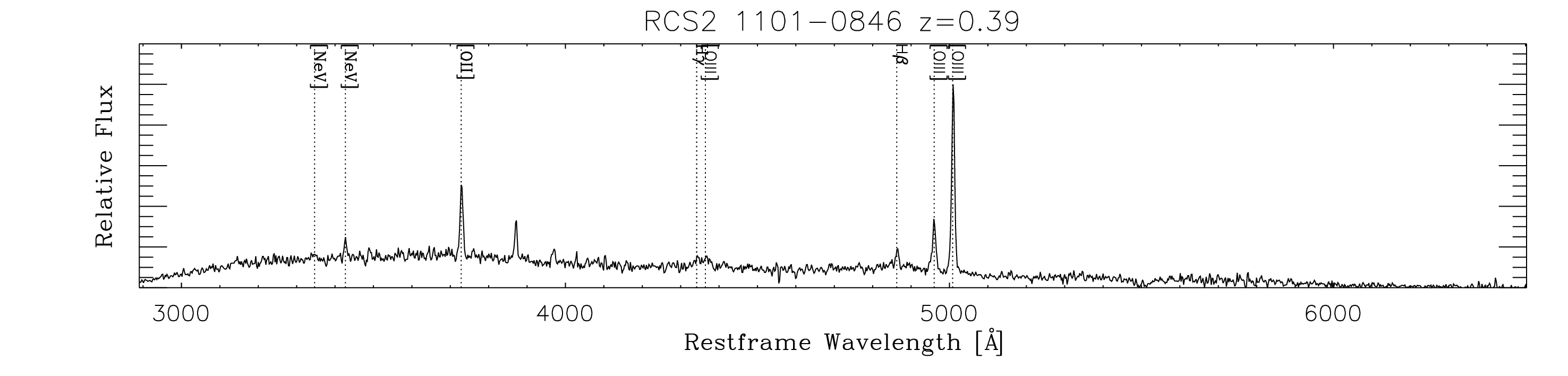}
\includegraphics[scale=0.33,trim = 15mm 15mm 0mm 0mm, clip]{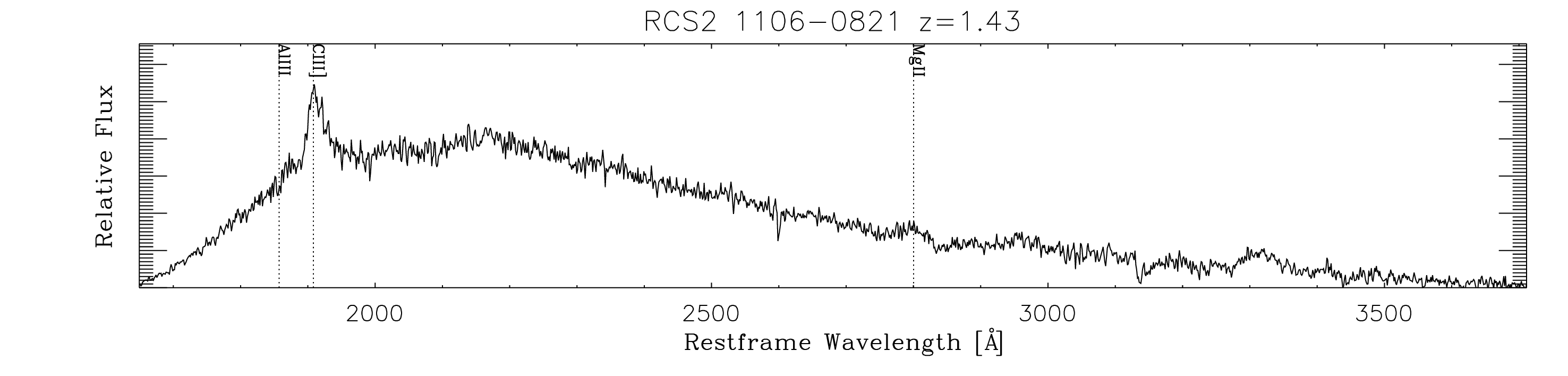}
\includegraphics[scale=0.33,trim = 15mm 0mm 0mm 0mm, clip]{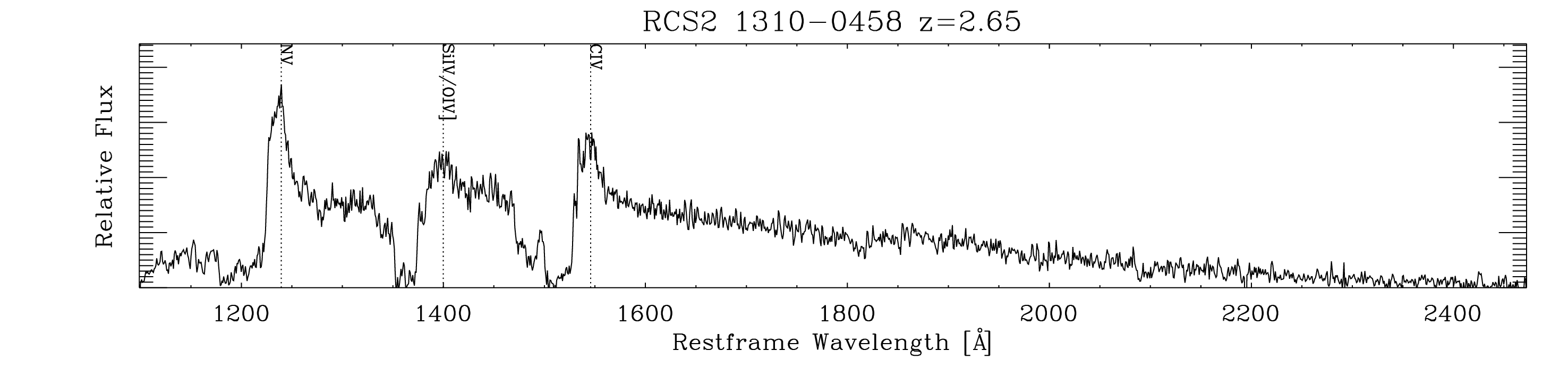}
\end{figure*}

\begin{acknowledgements}
\textbf{We thank the referee for useful comments and suggestions}. Support for L.F. Barrientos, K. Pichara and T. Anguita is provided by the Ministry of Economy, Development, and Tourism's Millennium Science Initiative through grant IC120009, awarded to The Millennium Institute of Astrophysics, MAS. We also acknowledge financial support from Proyecto Financiamiento Basal PFB06, Gemini Conicyt grant 32110010, Programa de Postgrado Instituto de Astrof\'isica and from BASAL PFB-06, and FONDEF D11I1060. L. F. Barrientos' research is supported by proyecto FONDECYT 1120676. T. Anguita acknowledges support by proyecto FONDECYT 11130630. D.N.A Murphy acknowledges support through FONDECYT grant 3120214.  

\end{acknowledgements}

%-------------------------------------------------------------------

\bibliographystyle{aa}
\bibliography{biblio}

\end{document}